\documentclass[11pt,a4paper]{article}

\usepackage[english]{babel}
\usepackage[utf8]{inputenc}
\usepackage{amsmath}
\usepackage{amsfonts}
\usepackage{graphicx}
\usepackage{epstopdf}
\usepackage{enumitem}
\usepackage[font=footnotesize, labelfont=bf]{caption}
\usepackage[figbotcap, loose]{subfigure}
\usepackage{mathptmx}       
\usepackage{bm}
\usepackage{siunitx}
\usepackage{tabularx}
\usepackage{booktabs}
\usepackage{multirow}
\usepackage{hyperref}
\usepackage{setspace}

\usepackage[square,numbers,sort&compress]{natbib}

\def\vec#1{\boldsymbol{#1}}
\usepackage{placeins}

\usepackage{chngcntr}
\counterwithin*{equation}{section}

\usepackage{titlesec}                     
\titleformat{\paragraph}
{\normalfont\normalsize\bfseries}{\theparagraph}{1em}{}

\DeclareMathAlphabet{\mathcal}{OMS}{cmsy}{m}{n}

\usepackage[symbol]{footmisc}

\begin{document}

\doublespacing
\noindent\textbf{\LARGE Quantitative \textit{in vivo} imaging to enable tumor forecasting and treatment optimization}
\singlespace

\vspace{2mm}

{\large
\noindent Guillermo Lorenzo$^{1,2,}$\footnote[1]{\textbf{Corresponding author:} guillermo.lorenzo@unipv.it, guillermo.lorenzo@utexas.edu}, David A. Hormuth II$^{2,3}$, Angela M. Jarrett$^{2,3}$, Ernesto A. B. F. Lima$^{2,4}$, Shashank Subramanian$^2$, George Biros$^2$, J. Tinsley Oden$^2$, Thomas J. R. Hughes$^2$, and Thomas E. Yankeelov$^{2,3,5,6}$
}

\vspace{2mm}

{\small \noindent
$^1$Department of Civil Engineering and Architecture, University of Pavia, via Ferrata 3, 27100 Pavia, Italy\\
$^2$Oden Institute for Computational Engineering and Sciences, The University of Texas at Austin,  201 E. 24th St., Austin, TX 78712-1229, USA\\ 
$^3$Livestrong Cancer Institutes, Dell Medical School,  The University of Texas at Austin, 1601 Trinity St., Bldg. B, Austin, TX 78701, USA\\
$^4$Texas Advanced Computing Center, The University of Texas at Austin, J.J. Pickle Research Campus, Bldg. 205, 10100 Burnet Rd. (R8700), Austin, TX 78758, USA\\
$^5$Departments of Biomedical Engineering, Diagnostic Medicine, and Oncology, The University of Texas at Austin, 107 W. Dean Keeton St., Austin, TX 78712, USA\\
$^6$Department of Imaging Physics, MD Anderson Cancer Center, 1515 Holcombe Blvd., Houston, TX 77030, USA
}

\vspace{5mm}

{\noindent \large \textbf{Abstract}}

\noindent Current clinical decision-making in oncology relies on averages of large patient populations to both assess tumor status and treatment outcomes. However, cancers exhibit an inherent evolving heterogeneity that requires an individual approach based on rigorous and precise predictions of cancer growth and treatment response. To this end, we advocate the use of quantitative \emph{in vivo} imaging data to calibrate mathematical models for the personalized forecasting of tumor development. In this chapter, we summarize the main data types available from both common and emerging \emph{in vivo} medical imaging technologies, and how these data can be used to obtain patient-specific parameters for common mathematical models of cancer. We then outline computational methods designed to solve these models, thereby enabling their use for producing personalized tumor forecasts \emph{in silico}, which, ultimately, can be used to not only predict response, but also optimize treatment. Finally, we discuss the main barriers to making the above paradigm a clinical reality.

\vspace{2mm}


\section{Introduction to tumor forecasting}\label{sec:intro}

Cancers are highly \emph{heterogeneous} diseases supported by diverse biological mechanisms occurring, interacting, and evolving at multiple spatial and temporal scales \cite{Marusyk2010}. These phenomena span from the phenotypic and genotypic cellular diversity within the tumor to the regional variations of the tumor microenvironment (e.g., vasculature and extracellular matrix), which can result in epigenetic changes in cancer cells or gradients in nutrient availability. Hence, the heterogeneous nature of cancer makes each patient's case unique. 
However, established, standard-of-care methods determine diagnosis, stage, treatment regimen, and response to treatment according to historical population averages.
This paradigm only enables the observation of cancer evolution and the outcome of treatment at fixed time points, offers a limited individualization of disease management, and largely ignores the intrinsic heterogeneity of cancers, which may result in treatment failure \cite{Jain1998,Withers1985}.
Thus, a new clinical paradigm that effectively integrates the spatiotemporal dynamics of tumor growth and treatment response to identify effective clinical strategies for each patient is desperately needed.  We posit that mathematical modeling informed by clinically-relevant data can provide the framework to address this challenge \cite{Yankeelov2013,Yankeelov2015,Rockne2019}.

\emph{Computational oncology} is a rapidly growing field that attempts to leverage mathematical models of the key biological mechanisms that characterize cancer to predict how  a patient's tumor will grow and respond to treatment  \cite{Rockne2019,Yankeelov2013,Yankeelov2015}. 
Computer simulations of these models provide personalized \emph{tumor forecasts}, designed to ultimately assist oncologists in clinical decision-making. 
For example, tumor forecasts may predict disease progression, thereby providing much needed guidance on the optimal intervention strategy early in the course of therapy.
Indeed, we hypothesize that treatment optimization can be achieved through the development and rigorous validation of practical mathematical models and efficient computational methods that can provide accurate personalized predictions of cancer development and treatment response.

A fundamental challenge in computational oncology is accomplishing the \emph{patient-specific parameterization} of the biological mechanisms involved in cancer models (e.g.,  tumor cell mobility, proliferation and death rates, or therapy efficacy). In general, these parameters are extremely difficult to measure \emph{in vivo} in human tumors. 
However, medical imaging may provide a viable source of data for this purpose.
Clinical oncology currently focuses on \emph{anatomical imaging} for the diagnosis, treatment, monitoring, and assessment of therapeutic response of solid tumors \cite{Therasse2000} (e.g., measuring tumor size, identifying invasion into adjacent structures, and detecting metastasis). 
Unfortunately, anatomical imaging frequently fails to capture the heterogeneous underlying biology within tumors. 
Alternatively, \emph{quantitative imaging} techniques enable the measurement of clinically-relevant biological features of tumors, such as tumor cell density, blood volume fraction, and perfusion \cite{HormuthII2019}.
Thus, these quantitative imaging techniques can be used to assess the spatiotemporal evolution of a cancer's heterogeneous architecture, morphology, growth dynamics, and response to therapy, thereby providing the necessary data to parameterize predictive models of tumor growth and treatment \cite{Yankeelov2013,HormuthII2019a}.

In this chapter, we will discuss how quantitative imaging can be used to enable tumor forecasting and optimization of therapeutic response. We will begin by identifying relevant quantitative imaging data types and how they are incorporated into existing image-based models of cancer growth and treatment. We will also provide insights into the technical aspects of numerical implementation, model calibration, and model selection.  Then, we will introduce a promising framework to optimize patient treatment plans. We will conclude with a discussion of the barriers to successfully translating image-based computational tumor forecasting into patient care.

\section{Relevant data types from medical imaging}\label{sec:img}

While measuring tumor size throughout therapy is central in oncological response assessments \cite{Therasse2000}, the dynamics of tumor size changes are often temporally downstream of intratumoral biological and physiological responses to therapy.
Magnetic resonance imaging (MRI) and positron emission tomography (PET) provide non-destructive and non-invasive 3D quantitative measurements of biological properties within and around the tumor. Hence, the acquisition of these imaging data at several timepoints is well-suited to initialize and parameterize mathematical models of tumor growth and treatment response.
In this section, we will briefly introduce the relevant MRI and PET measurements that have been commonly used in computational oncology (see Fig.~\ref{fig:img} for representative images of these techniques). For a detailed review of advanced MRI and PET techniques in oncology, the  reader is respectively referred to \cite{HormuthII2019} and \cite{Gambhir2002}.

\subsection{Diffusion weighted magnetic resonance imaging}\label{sec:dwi}

Diffusion weighted (DW-) MRI is an established technique that has been applied in oncology as a noninvasive assessment of cellularity changes during treatment \cite{Padhani2009}. DW-MRI is sensitive to the diffusion of water molecules within tissue. 
In a DW-MRI experiment, water molecules are first tagged based on their spatial location. Then, after a short delay of typically 20–-60 ms, a second spatial-encoded tag is applied. During this delay, water molecules move throughout the tissue due to diffusion. If the water molecules do not travel far, the first spatial-encoded tag can be largely removed by the second spatial-encoded tag and there is no loss (or gain) in signal intensity. However, if the water molecules move throughout the domain, there is a net-difference between the two spatial-encoded tags resulting in a decrease in signal intensity. Thus, the signal intensities within each voxel in the resulting image are ``weighted'' based on water diffusion. In practice, several diffusion weighted experiments are performed with different settings (e.g., varied diffusion-sensitizing gradient amplitudes of the magnetic field) to spatially quantify the apparent diffusion coefficient ($ADC$) of water.  However, water diffusion in tissue is heavily restricted by cells, macromolecules, and extracellular structures. Hence, these physical barriers reduce the measured $ADC$. This phenomenon has been observed in several studies showing an inverse correlation between $ADC$ and cellularity \cite{Anderson2000,Jiang2016,Barnes2015}. Following these reports' results, $ADC$ can be used to estimate cellularity using:
\begin{equation}\label{eq:NADC}
N(\vec{x},t) = \theta\left( \frac{ADC_w - ADC(\vec{x},t)}{ADC_w - ADC_{\min}}\right)
\end{equation}
where $\theta$ represents the maximum tumor cell carrying capacity for an imaging voxel (determined by the voxel dimensions and assumptions in cell geometry and packing density), $ADC_w$ is the $ADC$ of free water at 37$^{\circ}$C (i.e., $2.5\cdot 10^{-3}$ mm$^2$/s; \cite{Whisenant2014}), $ADC(\vec{x},t)$ is the $ADC$ value at a given 3D position $\vec{x}$ and time $t$, and $ADC_{\min}$ is the minimum $ADC$ value observed within the tumor. Fig.~\ref{fig:img} displays a representative $ADC$ map from breast and brain cancer.
While there are significant correlations between cellularity and the measured $ADC$, cellularity is not the sole factor in dynamic changes in $ADC$. Changes in cell size, cell permeability, and tissue tortuosity may alter the measured $ADC$ \cite{Padhani2009}. Other diffusion-based imaging approaches can also report on cell size \cite{Jiang2016} and diffusion anisotropy \cite{Sundgren2004}. The reader is referred to \cite{Koh2007} for a technical review of DW-MRI and its applications in oncology.

\subsection{Dynamic contrast-enhanced magnetic resonance imaging}\label{sec:dce}
\begin{figure}[t]
\centering
\includegraphics[width=\textwidth]{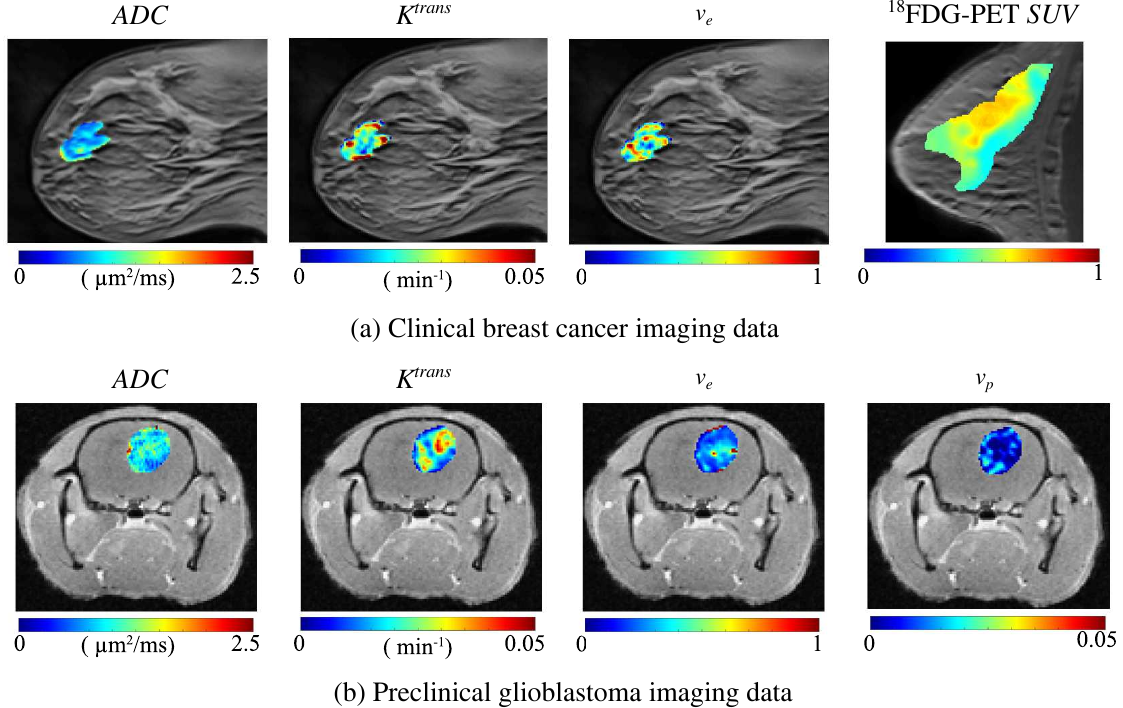}
\caption{Representative quantitative imaging measurements from clinical and preclinical settings. (a) Parameter maps extracted from DW-MRI, DCE-MRI, and $^{18}$FDG-PET through the central slice of a breast tumor (all MR data are from the same patient, PET data are from a different individual).
DW-MRI provides estimates of $ADC$, while DCE-MRI yields estimates of $K^{trans}$ and $v_e$. The $^{18}$FDG-PET $SUV$ map shows increased glucose uptake within the breast tumor relative to surrounding tissue. (b) Parameter maps acquired in a preclinical murine model of glioblastoma from DW-MRI ($ADC$) and DCE-MRI ($K^{trans}$, $v_e$, and $v_p$).  }
\label{fig:img}
\end{figure}

Dynamic contrast-enhanced (DCE-) MRI consists of the rapid acquisition of a series of heavily $T_1$-weighted images before, during, and after the injection of a $T_1$ altering-contrast agent (typically a Gadolinium chelate) to probe vascular properties in tissue \cite{Yankeelov2009}. Using a pre-contrast $T_1$ map, any post-contrast $T_1$ changes can be  related to the concentration of the contrast agent. Thus, each image voxel yields a signal intensity time course that can be related to the concentration of the contrast agent within that voxel. The subtraction images obtained from pre- and post-contrast enhanced images are often used to identify tumor regions, which usually show areas of rapid and intense enhancement due to their higher and more permeable vascularity than the neighboring healthy tissue. The dynamics of signal intensity are commonly analyzed with a two-compartment pharmacokinetic model describing the extravasation of the contrast agent from the plasma space to the tissue space \cite{Yankeelov2009}. The solution to this model  is given by 
\begin{equation}\label{eq:dce}
C_t(\vec{x},t) = K^{trans}(\vec{x})\int_0^t C_p(u)e^{-\frac{K^{trans}(\vec{x})}{v_e(\vec{x})}(t-u)}du + v_p(\vec{x})C_p(t),
\end{equation}
\noindent where $C_t(\vec{x},t)$ is the concentration of the contrast agent in tissue at position $\vec{x}$ and time $t$, $C_p(t)$ is the concentration of the contrast agent in the plasma space at time $t$, $ K^{trans}(\vec{x})$ is the volume transfer constant from the plasma to tissue space, $v_e(\vec{x})$ is the extravascular-extracellular volume fraction, and $v_p(\vec{x})$ is the plasma volume fraction. Importantly, $C_t$, $K^{trans}$, $v_e,$ and $v_p$ are all voxel-specific and are related to structural (cell density) and physiological (vessel permeability and perfusion) properties. $C_p(t)$ can be measured directly for individual subjects from a large artery within the image field of view or can be replaced with a population-based estimate \cite{Li2011}. Fig.~\ref{fig:img} shows $K^{trans}$, $v_e,$ and $v_p$ maps from a preclinical and a clinical study. 

\subsection{Molecular imaging with positron emission tomography}\label{sec:pet}
PET relies on the injection of a radiopharmaceutical (or PET tracer) to generate image contrast.  As there is no endogenous signal, PET has excellent sensitivity to detect and localize the distribution of radiopharmaceuticals throughout the body.  Several radiopharmaceuticals have been developed to probe tumor properties, such as glucose metabolism (\emph{via} $^{18}$F-fludeoxyglucose  or $^{18}$FDG), hypoxia (\emph{via} $^{18}$F-fluoromisonidazole or $^{18}$F-MISO), cellular proliferation (\emph{via} $^{18}$F-Flurodeoxythymidine \cite{Woolf2014}), and receptor status (e.g., $^{64}$Cu-diethylenetriaminepen-taacetic acid Trastuzumab for HER2+ positive cancers \cite{Mortimer2018}). We will primarily focus on $^{18}$FDG and $^{18}$F-MISO as they are well-established in oncology, but the analysis techniques are similar for other PET tracers. 
  
In an $^{18}$FDG-PET study, a single image is acquired following the injection of the glucose analogue $^{18}$FDG, which cells uptake in a similar fashion to glucose.
However, once internalized, $^{18}$FDG is phosphorylated and trapped intracellularly. The resulting image intensities are proportional to the concentration of $^{18}$FDG within each voxel. PET tracer uptake can be quantified using the standardized uptake value ($SUV$), which is the ratio of the concentration of $^{18}$FDG radioactivity in tumor tissue to the total injected dose and divided by the patient's body weight. In oncology studies, contrast between tissues is typically generated due to variations in glucose uptake due to an overexpression of glucose transporters and hexokinase activity in tumor cells relative to healthy cells \cite{Castell2008}. This difference in $^{18}$FDG uptake is also shown in Fig.~\ref{fig:img}.

Likewise, in $^{18}$F-MISO PET a single image is also acquired following the injection of  $^{18}$F-MISO, which is a radiopharmaceutical that produces images sensitive to oxygen concentration in tissue \cite{Rajendran2015}. After $^{18}$F-MISO is internalized by cells, it is reduced to produce a radical anion. In normoxic or oxygen-rich environments, oxygen accepts the electron from the radical anion enabling $^{18}$F-MISO to leave the cell. Conversely, in hypoxic or oxygen-poor environments, the radical anion of $^{18}$F-MISO binds to other intracellular macromolecules trapping it within the cell. Thus, the concentration of $^{18}$F-MISO and the produced PET signal within a voxel are inversely proportional to the oxygen concentration. $^{18}$F-MISO uptake is quantified using the standardized uptake value ($SUV$) or the oxygen enhancement ratio ($OER$), which is the ratio of signal intensity in tumor relative to blood.

\section{Image-based mathematical models of cancer}\label{sec:cmodels}

Medical imaging provides an excellent way to develop, calibrate, and validate personalized mathematical models of cancer evolution and treatment response \cite{Yankeelov2013} for three main reasons.
First, medical imaging enables the \emph{in vivo} measurement of relevant biological properties in tumor and healthy tissues, which would otherwise be impractical or impossible to measure in individual patients. 
Second, medical imaging data can be obtained frequently throughout the clinical management of the patient's tumor, which enables model calibration. 
Third, medical imaging data are acquired on a regular voxel grid, which facilitates their computational processing.
In this section, we discuss common image-based models of tumor growth and treatment response that leverage the quantitative imaging data types introduced in Section~\ref{sec:img}. Fig.~\ref{fig:models} shows simulation outputs of many of the models discussed in this section. 

\subsection{Baseline tumor growth models}\label{sec:base}  
In mathematical oncology \cite{Rockne2019}, the \emph{logistic growth} model is one of the simplest and most common approaches to describe changes in tumor volume \cite{Benzekry2014} or cell number \cite{Atuegwu2011} over time. It is a flexible model that can be adapted to \textsl{in vitro} and \textsl{in vivo} data alike. The formulation of the logistic growth model over a certain tissue region of interest follows the partial differential equation (PDE):
\begin{equation}\label{eq:log}
\frac{\partial N(\vec{x},t)}{\partial t}=k(\vec{x})N(\vec{x},t)\left(1-\frac{N(\vec{x},t)}{\theta}\right),
\end{equation}
where $N(\vec{x},t)$ is the \emph{tumor cell density} at position $\vec{x}$ and time $t$, $k(\vec{x})$ is a spatially-varying net proliferation rate, and $\theta$ is the carrying capacity. 
The image-informed applications of this model have typically been posed voxelwise, such that $N(\vec{x},t)$ is redefined as the number of cancer cells within the voxel in position $\vec{x}$ at time $t$.
Atuegwu \textsl{et al.} \cite{Atuegwu2011} used this approach to predict tumor growth in breast cancer patients receiving neoadjuvant chemotherapy.
First, they used Eq.~\eqref{eq:NADC} to estimate $N(\vec{x},t)$ from $ADC$ maps obtained \emph{via} DW-MRI. The estimates of $N$ at baseline (pre-treatment) and after one cycle of therapy were used to determine $k(\vec{x})$.
Then, Atuegwu \textsl{et al.}  used  their model equipped with the resulting $k(\vec{x})$  to predict $N$ at the conclusion of therapy. They observed a strong correlation between the predictions and data estimates of $N$ over the entire tumor (Pearson correlation coefficient, PCC, of 0.95) and for individual voxels (PCC=0.70). Fig.~\ref{fig:models} shows an example of this approach for a clinical breast cancer model and a preclinical glioblastoma model.

However, the logistic growth model fails to capture the potential movement of cells that may occur over time. To overcome this limitation, the logistic growth model can be extended to a \emph{reaction-diffusion} model given by the PDE
\begin{equation}\label{eq:rd}
\frac{\partial N(\vec{x},t)}{\partial t} = \nabla\cdot\left(D \nabla N(\vec{x},t)\right) + k(\vec{x})N(\vec{x},t)\left(1-\frac{N(\vec{x},t)}{\theta}\right),
\end{equation}
where the first term on the right is a diffusion term describing the movement of tumor cells with a constant diffusion coefficient $D$, while the second term on the right is a reaction term represented by the logistic growth of cancer cells. 

Eq.~\eqref{eq:rd} is well established in computational oncology  \cite{Hormuth2019,Weis2013,Jarrett2018a,Rockne2015,Swan2018,Hogea:2008b,Wong:2017a,Roque2018}.
The work of Swanson \textsl{et al.} \cite{Baldock2013} in high grade gliomas showed one of its first image-informed applications, using anatomical MRI data to provide segmentation of tumor boundaries and fixed cell-density counts in enhancing and non-enhancing disease. The spatiotemporal changes in tumor boundaries were used to estimate a constant tumor-specific proliferation rate $k$ and tissue-specific $D$, which were then used to predict patient survival. This approach has had promising results in relating growth kinetics to patient outcomes \cite{Baldock2013,Neal2013}. However, it does not describe the intratumoral heterogeneity of cell density. 
Hormuth \textsl{et al.} \cite{Hormuth:2015a} addressed this limitation by estimating $N$ from $ADC$ maps obtained \emph{via} DW-MRI using Eq.~\eqref{eq:NADC} in a murine model of glioma. In this study, animals were imaged up to seven times over ten days. The first three imaging datasets were used to initialize $N$ as well as to calibrate $k(\vec{x})$ and $D$. In a separate calibration, a spatially-constant $k$ (i.e., tumor specific) was also calibrated along with $D$. The calibrated model was then used to predict $N(\vec{x},t)$ at the remaining imaging visits. While both calibration scenarios overestimated future tumor growth, the predictions with a spatially-varying $k(\vec{x})$ rendered lower tumor volume errors, higher Dice correlation coefficients, and higher concordance correlation coefficients (CCC; all $p<0.05$). These results highlight the importance of accounting for the intratumoral heterogeneous dynamics to obtain accurate tumor forecasts and the promising potential of quantitative imaging to provide the required data for this purpose. Other studies have used alternative quantitative imaging measures to inform the reaction-diffusion model in Eq.~\eqref{eq:rd}; examples include incorporating anisotropic diffusion \emph{via} diffusion tensor imaging (DTI, a form of DW-MRI \cite{Sundgren2004}) \cite{Swan2018,Jbabdi2005}, using cell density measurements \emph{via} contrast-enhanced computed tomography \cite{Wong:2017a}, and estimating cell phenotypes \emph{via} DCE-MRI \cite{Roque2018}.

Alternatively, \emph{phase-field models} are another common  paradigm to describe tumor growth \cite{Lima2016,Agosti2018,Lorenzo2016,Lorenzo2019,Xu2020}. The \emph{phase field} $\phi(\vec{x},t)$  identifies healthy tissue (e.g., $\phi=0$) from tumor tissue (e.g., $\phi=1$), showing a smooth and thin transition between either region. Phase-field models usually focus on the dynamics of tumor morphology through the evolution of the healthy-tumor interface, which is implicitly defined by a phase-field isosurface.
These models rely on more complex physics than those described above, usually requiring the definition of an energy functional that drives tumor growth  \cite{Gomez2017,Lima2016,lima17}.  
Phase-field modeling has been notably applied in brain tumors \cite{Lima2016,Agosti2018}, prostate cancer \cite{Lorenzo2016,Lorenzo2019}, and tumor angiogenesis \cite{Xu2020,Vilanova2017}.
However, there is a paucity of studies using quantitative imaging data to inform phase-field models. This is partly due to their more complex dynamics, which usually requires a higher number of parameters, larger and richer patient-specific datasets, more advanced numerical methods, and more computational resources.

We identified two illustrative works that use imaging measurements to initialize, calibrate, and/or validate phase-field models of cancer. 
Lima \textsl{et al.} \cite{Lima2016,lima17} have been investigating model selection to identify the best formulation of murine glioma growth according to longitudinal anatomical MRI tumor measurements, including an array of phase-field models also accounting for the local tumor-induced mechanical stress field (see Section~\ref{sec:mech}). Their work shows that phase-field models are plausible formulations of tumor growth and in \cite{Lima2016} they emerge as the best models indeed.
Additionally, Agosti \textsl{et al.} \cite{Agosti2018} developed a  phase-field model of glioblastoma multiforme that uses quantitative DTI data to define anisotropic tumor cell motility and nutrient diffusion. Their work focuses on the prediction of tumor recurrence after surgical resection and subsequent radiotherapy. By accounting for post-surgery changes in tissue architecture, they obtained a Jaccard index of 0.71 post-radiotherapy.

\subsection{Mechanically-coupled models}\label{sec:mech}

Local mechanical tissue properties and tumor-induced mechanical stresses are known to affect cancer growth dynamics \cite{Helmlinger1997,Nagelkerke2015,Jain2014}. 
For example, Helmlinger \textsl{et al.} \cite{Helmlinger1997} observed that tumor spheroid growth \textsl{in vitro} was increasingly inhibited as the substrate matrix stiffness was augmented. Uncontrolled tumor growth can also severely deform healthy tissue structures, thereby adversely impacting patient health and quality of life. Therefore, several mathematical models of cancer couple tumor growth dynamics with local mechanical equilibrium \cite{Lorenzo2019,Jarrett2018a,Weis2013,Hogea:2008b,HormuthII2017,Liu2014,Clatz2005}.

\begin{figure}[!t]
\centering
\includegraphics[width=\textwidth]{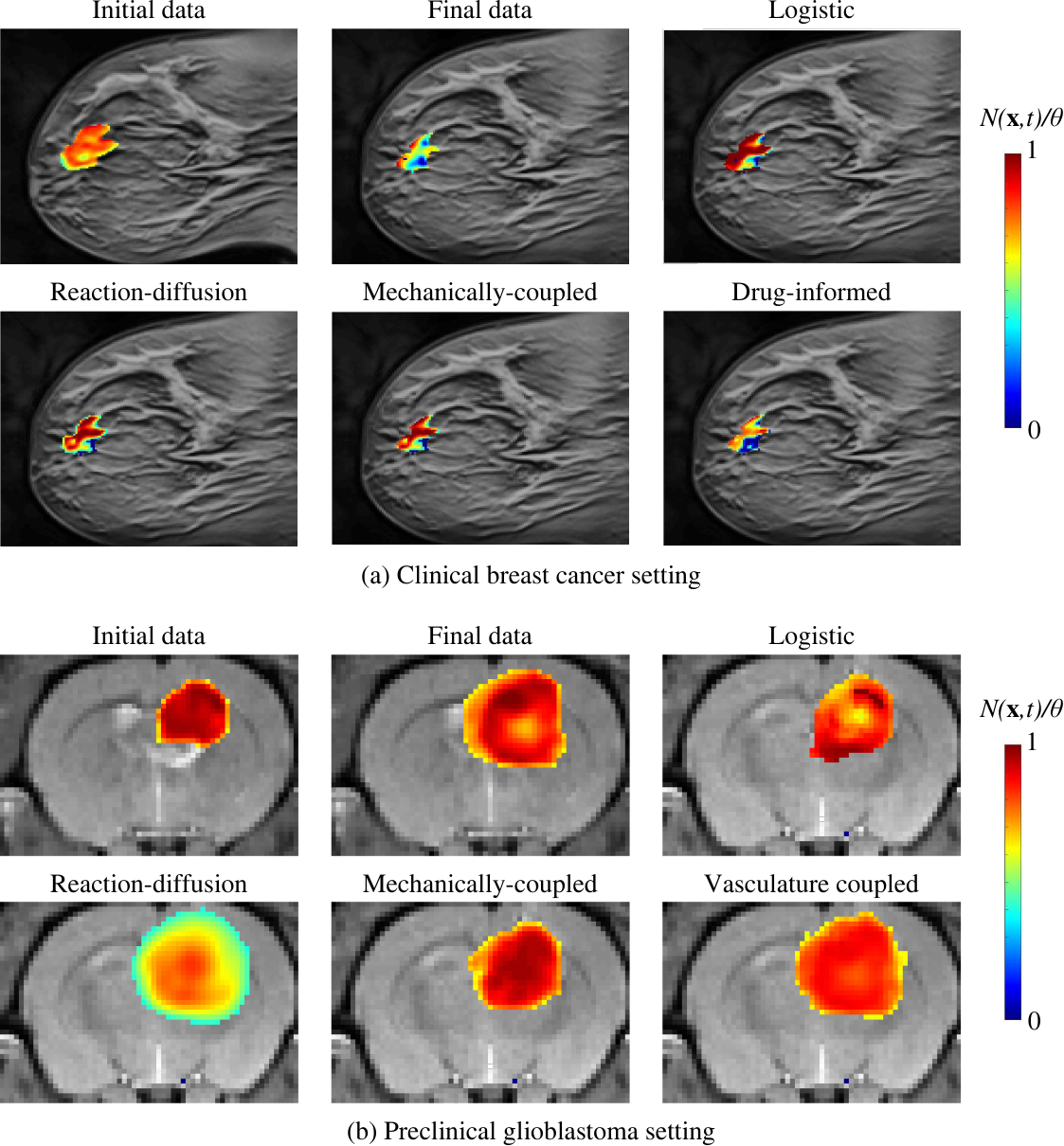}
\caption{Differences between image-based models in clinical and preclinical settings. (a) Example from a clinical breast cancer model, where measured initial, measured final, and model forecasts of the final distributions of tumor cell density are shown. The logistic model fails to capture the expansion of the tumor into nearby tissue, while the remaining models incorporating a diffusion term perform better. (b) Example from a preclinical model of glioblastoma, where measured initial, measured final, and model predicted final distributions of tumor cell density are shown. The logistic model also fails to predict the expansion, but does predict an intratumoral low cell density area.}
\label{fig:models}
\end{figure}

A common approach \cite{Weis2013} is to dampen the diffusion coefficient in Eq.~\eqref{eq:rd} with a function of local tissue stress: 
\begin{equation}\label{eq:vm}
D(\vec{x},t)=D_0e^{-\gamma_v\sigma_{vm}(\vec{x},t)},
\end{equation}
where $D(\vec{x},t)$ is now a spatially and temporally varying diffusion coefficient, $D_0$ is the tumor cell diffusion coefficient in the absence of stress, $\gamma_v$ is an empirical coupling constant, and $\sigma_{vm}(\vec{x},t)$ is the von Mises stress. Here, $\sigma_{vm}(\vec{x},t)$ is used to summarize the local mechanical stress field, which is calculated assuming quasistatic linear elastic equilibrium with tissue-specific mechanical properties:
\begin{equation}\label{eq:lineq}
\nabla\cdot\left(\lambda\left(\nabla\cdot\vec{u}\right)\vec{I} + \mu\left(\nabla\vec{u}+\nabla\vec{u}^T\right)\right) -  \gamma_N\nabla N = \vec{0},
\end{equation}
where $\lambda$ and $\mu$ are the Lam\'{e} coefficients (related to the tissue's Young modulus $E$ and Poisson's ratio $\nu$), $\vec{u}$ is the displacement field due to tumor cell growth, and $\gamma_N$ is another empirical coupling constant. In Eq.~\eqref{eq:vm}, the first term on the left-hand side represents the linear elastic tissue response to the local tumor-induced forces described by the second term on the left.
Weis \textsl{et al.} \cite{Weis2013} used tumor cell number estimates from DW-MRI data using Eq.~\eqref{eq:NADC} to initialize and calibrate a mechanically-coupled reaction-diffusion model of breast cancer growth during neoadjuvant chemotherapy consisting of Eqs.~\eqref{eq:rd}--\eqref{eq:lineq}. Their work shows that the mechanically-coupled model rendered more accurate predictions of $N$ (PCC=0.85) than the baseline reaction-diffusion model (PCC=-0.29). Several subsequent studies have also used Eqs.~\eqref{eq:vm}--\eqref{eq:lineq} to couple mechanics to breast and brain tumor dynamics \cite{Jarrett2018a,HormuthII2017}, as shown in Fig.~\ref{fig:models}. Lima \emph{et al.} further considered a mechanical inhibition of tumor proliferation following a similar formulation to Eq.~\eqref{eq:vm} \cite{Lima2016,lima17}. Moreover, the prostate cancer model of Lorenzo \emph{et al.} \cite{Lorenzo2019} extended Eq.~\eqref{eq:vm} to combine the measure of mechanical tissue distortion \emph{via}  $\sigma_{vm}$ with hydrostatic stress, which is not captured by $\sigma_{vm}$ and contributes to a more precise description of intratumoral stress. 

Other mathematical models couple local mechanics to tumor growth dynamics through a reaction-advection-diffusion equation \cite{Hogea:2008b,Liu2014} in which the tumor cell drift velocity is related to the displacement field, thereby explicitly simulating the displacement of cells due to mechanical deformation.
Additionally, Wong \textsl{et al.} \cite{Wong:2017a} leveraged a hyperelastic biomechanical model. Interestingly, in this study tumor dynamics was described using a reaction-diffusion model in which the proliferation rate $k(\vec{x})$ was calibrated with $^{18}$FDG-PET $SUV$ data, as follows:
\begin{equation}\label{eq:ksuv}
k(\vec{x}) = \frac{\alpha SUV(\vec{x}) - \beta c(\vec{x},t)}{c(\vec{x},t)\left(1-c(\vec{x},t)\right)},
\end{equation}
where $SUV(\vec{x})$ is the standardized uptake value at position $\vec{x}$, $\alpha$ and $\beta$ are unknown constants to be calibrated, and $c(\vec{x},t)$ is the cell volume fraction estimated from computed tomography at position $\vec{x}$ and time $t$.

\subsection{Vasculature-coupled models}\label{sec:vasc}

Co-opting of local vasculature and recruitment of new blood vessels \emph{via} angiogenesis is a critical component of cancer development that is needed to support growth past 2--3 mm$^3$ in size \cite{Gillies1999,Jain2007}. 
Thus, understanding the evolving distribution and function of the tumor-supporting vasculature is crucial to accurately model tumor growth and treatment response. 
There is an extensive literature on mathematical models of tumor angiogenesis  \cite{Vilanova2017}. However, very few describe this phenomenon at the imaging/tissue scale \cite{Hormuth2019,Swanson2011} or personalize it for individual tumors. 

Hormuth \textsl{et al.} \cite{Hormuth2019} developed a murine model of glioma growth coupled with angiogenesis that was initialized and calibrated using tumor cell number estimates obtained from DW-MRI \emph{via} Eq.~\eqref{eq:NADC} and estimates of the blood volume fraction extracted from DCE-MRI. 
The spatiotemporal evolution of tumor cells and vasculature was described using two coupled reaction-diffusion equations. 
In this model, vasculature influenced the direction of tumor growth and was coupled to the carrying capacity. Similarly, tumor cells also influenced the direction of vasculature  evolution. The animals were imaged up to seven times over a period of ten days. Model parameters were calibrated using the first three imaging datasets, and then used in a forward evaluation of the model to predict tumor growth at the remaining imaging time points. The authors observed that their model resulted in less than 10.3$\%$ error in tumor volume predictions and less than 9.4$\%$ error at the voxel-level for all prediction time points. Fig.~\ref{fig:models} shows an example of this approach in a pre-clinical model of glioblastoma.

Roque \textsl{et al.} \cite{Roque2018} developed a vasculature-informed preclinical reaction-diffusion model of breast cancer accounting for normoxic, hypoxic, and necrotic cancer subpopulations along with nutrient dynamics, which regulates normoxic cell proliferation as well as the normoxic-hypoxic and hypoxic-necrotic transfer rates. While not explicitly evolving the tumor-supporting vascular network, the authors used vasculature-derived parameters obtained from DCE-MRI (e.g., blood flow, mean transit time, and maximum enhancement) to initialize all model variables and calibrate key parameters. While the study results suggested that further model development is needed to capture individual differences in tumor growth, this work is a unique effort to identify tumor subpopulations using quantitative imaging data.

\subsection{Radiotherapy}\label{sec:rt}
Radiotherapy is a common treatment for many cancers \cite{Miller2016}. However, intratumoral heterogeneity may result in significant variations in treatment response, which may ultimately lead to poor therapeutic outcomes \cite{Gillies1999,Baumann2016}.
Image-based modeling could prove valuable to predict the response to radiotherapy and hence optimize treatment protocols for individual patients. 
To this end, several studies have investigated incorporating imaging measures from PET \cite{Rockne2015} and MRI \cite{Hormuth2020,Hormuth2018,Hathout2016,lima17} into reaction-diffusion based models to characterize patient response to radiotherapy. The usual approach to model radiotherapy effects is by instantaneously killing a fraction of tumor cells at treatment times \cite{Rockne2015,Hormuth2020,Hormuth2018,Hathout2016,lima17}. 
This strategy may be further combined with a transient or permanent reduction in tumor cell proliferation \cite{Hormuth2020,Hormuth2018,lima17}. These radiotherapy effects are usually modeled as a function of the prescribed dose, which may also account for local tumor cell and vascular densities.

Rockne \textsl{et al.} \cite{Rockne2015} adapted the glioblastoma model by Swanson \textsl{et al.} \cite{Baldock2013} to explicitly incorporate cell death due to radiotherapy based on $^{18}$F-MISO PET data. 
Oxygen concentration and the degree of hypoxia in tumors are known to significantly impact response to radiotherapy \cite{Vaupel2007}. 
Thus, Rockne \textsl{et al.} used $^{18}$F-MISO PET to assess the level of hypoxia by calculating the $OER$, which is then used along with the usual linear quadratic model of radiotherapy response \cite{Douglas1976} to calculate cell survival, $S$, as follows: 
\begin{equation}\label{eq:oer}
S=\exp\left( -\alpha\left(OER(\vec{x})\right)\left( d + \frac{d^2}{\alpha/\beta\left(OER(\vec{x})\right)} \right)\right),
\end{equation}
where $d$ is the prescribed radiation dose while $\alpha\left(OER(\vec{x})\right)$ and $\alpha/\beta\left(OER(\vec{x})\right)$ are radiosensitivity parameters as a function of the $OER$ at position $\vec{x}$.  
Rockne \textsl{et al.} observed that predictions by a model featuring Eq.~\eqref{eq:oer} outperformed those obtained with a model with uniform radiosensitivity (1.1$\%$ \emph{vs} 14.6$\%$ error in tumor volume, respectively). 

\subsection{Chemotherapy}\label{sec:ct}

Chemotherapy is another common treatment for most cancers \cite{Miller2016}. 
Unlike the localized nature of radiation therapy, chemotherapy relies on drugs that are administered systemically throughout the body. 
While chemotherapy has traditionally leveraged cytotoxic drugs (i.e., promoting cell death), recent approaches also use drugs targeting specific cancer cell markers to decrease proliferation or triggering particular immune responses. 
Similar to radiation therapy, challenges for modeling chemotherapies stem from quantifying how much drug is distributed in the tumor (and healthy tissues) and patient-specific treatment efficacy. Multiple mathematical models have been proposed to describe the effect of chemotherapy on tumor growth \cite{Swan1990,Yin2019,Swierniak2009}, but only a few are informed by quantitative imaging data. 

In particular, the contribution by Jarrett \emph{et al.}  \cite{Jarrett2018a} extended the mechanically-coupled reaction diffusion model consisting of Eqs.~\eqref{eq:rd}--\eqref{eq:lineq} by including the dynamic effect of chemotherapy in the tumor growth equation:
\begin{align}\label{eq:ct}
\frac{\partial N(\vec{x},t)}{\partial t} = \nabla\cdot\left(D(\vec{x},t) \nabla N(\vec{x},t)\right)  + k(\vec{x})N(\vec{x},t)\left(1-\frac{N(\vec{x},t)}{\theta}\right)  - \alpha C_d(\vec{x},t) N(\vec{x},t) 
\end{align}
where $\alpha$ is the patient-specific drug efficacy  and $C_d(\vec{x},t)$ is the drug concentration in the tissue.  $C_d(\vec{x},t)$ was approximated patient-wise by means of the two-compartment model commonly used to analyze the contrast agent pharmacokinetics in DCE-MRI data (see Section~\ref{sec:dce}). This approach has two central limitations: it assumes that the drug and the contrast agent have similar dynamics, and that all chemotherapies explicitly induce tumor cell death. However, Jarrett \emph{et al.} showed that their drug-informed model predictions outperformed those of the mechanically-coupled reaction-diffusion model without the drug term when compared to patient-specific estimates of tumor cell density extracted from DW-MRI \emph{via} Eq.~\eqref{eq:NADC} at the end of chemotherapy; in particular, the CCC improved from 0.85 to 0.99 ($p<0.01$).

\section{Computational methods to solve image-based cancer models}\label{sec:comp}
Mechanistic models of cancer usually consist of coupled, nonlinear PDEs. Using the appropriate numerical strategies, these cancer models can be solved and rendered as a computer simulation of the spatiotemporal development of a patient's tumor; i.e., a tumor growth forecast.
In this section we will provide an elementary description of the Finite Difference Method (FDM) \cite{Leveque2007}, Finite Element Analysis (FEA) \cite{Hughes2000} and Isogeometric Analysis (IGA) \cite{Cottrell2009}. All these numerical methods have been widely used to solve  PDEs  in science and engineering. 

\subsection{The Finite Difference Method}\label{sec:fdm}

\begin{figure}[t]
\includegraphics[width=\linewidth]{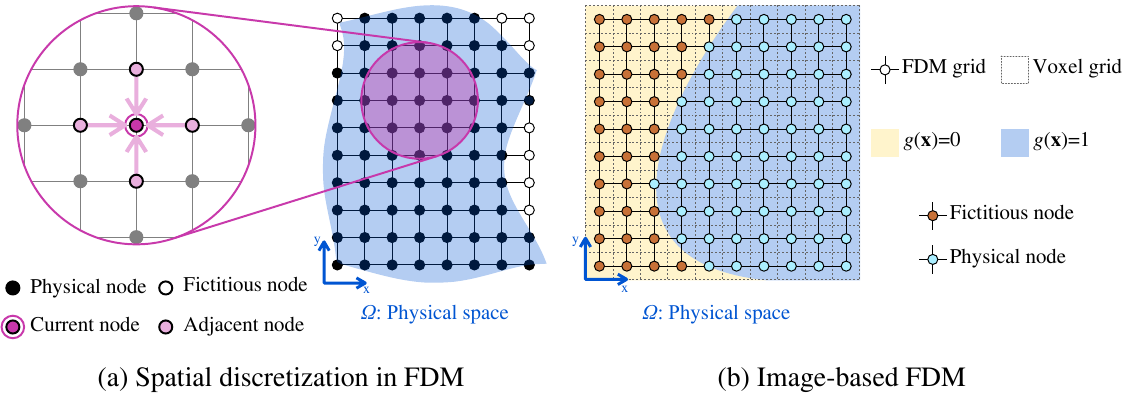}
\caption{The Finite Difference Method (FDM). 
(a) In FDM, the model equations are solved on a rectangular grid of physical nodes, which approximates the physical space $\Omega$ representing the problem's geometry. Ancillary fictitious nodes may also be required to construct the FDM grid. Spatial derivatives are approximated with linear combinations of the model solution at times $t_n$ and $t_{n+1}$ on each node and adjacent neighbors in each grid direction.
(b) The FDM can be formulated using imaging voxel data by defining one node per voxel. The organ segmentation can be used to build a map $g(\mathbf{x})$ to define $\Omega$ and hence identify physical ($g(\mathbf{x})=1$) and fictitious nodes ($g(\mathbf{x})=0$).}
\label{fig:fdm}
\end{figure}

The FDM relies on a direct approximation of the derivatives involved in the PDEs of the model by means of Taylor series expansions \cite{Leveque2007}. 
To apply the FDM, we define a global time interval for the simulation  $\left[0,T\right]$ and a geometric domain $\Omega$ consisting of a 3D box that includes the tumor-harboring organ. 
Let us discretize $\left[0,T\right]$ with a constant time step $\Delta t$, leading to a partition in time subintervals $[t_n, t_{n+1}]$, such that $t_{n+1}-t_n=\Delta t$, $t_0=0$, $t_{n_t}=T$, and  $n=0,\ldots,n_t-1$. 
We discretize $\Omega$ with a uniform 3D cartesian grid composed of $n_p=n_xn_yn_z$ nodes numbered $A=1,\ldots,n_p$, where $n_x$, $n_y$ and $n_z$ are the number of nodes in each spatial direction. Let $\vec{g}=(i,j,k)$ further denote the grid coordinates of each node, such that $i=0,\ldots,n_x-1$, $j=0,\ldots,n_y-1$, and $k=0,\ldots,n_z-1$. Then, the spatial coordinates of each node $A$ can be written as $\vec{x}_A=\vec{x}_0+\vec{h}\cdot\vec{g}$, where $\vec{h}=(h_x,h_y,h_z)$ is a vector holding the grid spacing in each spatial direction. Fig.~\ref{fig:fdm} illustrates an FDM grid in 2D. 

The standard FDM uses first-order approximations of the time and spatial derivatives in the PDE at the grid nodes and at a certain time instant $\tilde{t}\in[t_n, t_{n+1}]$. For example, the time derivative in Eq.~\eqref{eq:rd} would be approximated by
\begin{equation}
\frac{\partial N(\vec{x},t)}{\partial t} \approx \frac{ N(\vec{x}_A,t_{n+1}) - N(\vec{x}_A,t_{n})}{\Delta t} = \frac{ N^{n+1}_{i,j,k} - N^{n}_{i,j,k}}{\Delta t}
\end{equation}
\noindent on each node at $t=t_{n+1}$. Higher-order derivatives are recursively approximated with first-order approximations of the subsequent lower-order derivatives. 
Ultimately, the FDM method reduces the PDE on every node to an algebraic equation involving a combination of values of the PDE solution on the current and adjacent nodes in each spatial direction (see Fig.~\ref{fig:fdm}) at instants $t_n$ and $t_{n+1}$.
Then, the general strategy is to recursively use the known nodal values of the PDE solution at $t_n$, $\{N(\vec{x}_A,t_{n})\}_{A=1,\ldots,n_p}$, to calculate the nodal values at $t_{n+1}$, $\{N(\vec{x}_A,t_{n+1})\}_{A=1,\ldots,n_p}$.

Depending on the choice of $\tilde{t}$, there are three common FDM approaches in practice: the explicit Euler method ($\tilde{t}=t_n$), the implicit Euler method ($\tilde{t}=t_{n+1}$), and the Crank-Nicolson method ($\tilde{t}=t_n+\Delta t/2$) \cite{Leveque2007}. The explicit method enables the direct calculation of the PDE solution on the grid nodes at $t_{n+1}$,  $\{N(\vec{x}_A,t_{n+1})\}_{A=1,\ldots,n_p}$, from the nodal values of the solution at $t_n$. This approach involves a minimal computational cost, which has been exploited for the recursive model resolutions involved in the patient-specific calibration of image-based  models of brain and breast cancer \cite{Jarrett2018a,Hormuth2019}.
For example, the application of the explicit method to Eq.~\eqref{eq:rd} yields
\begin{align}\label{eq:explicit}
\frac{N^{n+1}_{i,j,k} - N^{n}_{i,j,k}}{\Delta t} & = D\left( \frac{  N^{n}_{i-1,j,k} - 2 N^{n}_{i,j,k} +  N^{n}_{i+1,j,k}}{h^2_x} + \frac{  N^{n}_{i,j-1,k} - 2 N^{n}_{i,j,k} +  N^{n}_{i,j+1,k}}{h^2_y}\right. + \notag \\ & + \left.\frac{  N^{n}_{i,j,k-1} - 2 N^{n}_{i,j,k} +  N^{n}_{i,j,k+1}}{h^2_z}\right) + k_{i,j,k}N^{n}_{i,j,k}\left(1-\frac{N^{n}_{i,j,k}}{\theta}\right)
\end{align}
\noindent at every grid node, where we have denoted $k(\vec{x}_A)=k_{i,j,k}$. Note that in Eq.~\eqref{eq:explicit}, we can directly compute $N^{n+1}_{i,j,k}$ from a linear combination of nodal values at $t_n$. However, the explicit method usually requires small time steps to ensure numerical stability. The implicit and Crank-Nicolson methods lead to a system of $n_p\times n_p$ algebraic equations whose resolution provides $\{N(\vec{x}_A,t_{n+1})\}_{A=1,\ldots,n_p}$. These FDM schemes are computationally more intensive, but show better numerical stability and enable the use of larger time steps. Application of these methods to nonlinear  PDEs like Eq.~\eqref{eq:rd} results in a nonlinear algebraic system, which can be solved with Newton's method by iteratively solving the corresponding linearized system \cite{Leveque2007}. Alternatively, an implicit-explicit method can use an implicit approach for the diffusion operator and an explicit scheme for the nonlinear logistic term  \cite{Leveque2007,Ruuth1995}, which leads to a linear algebraic system. Currently, multiple sparse-matrix algorithms enable a computationally efficient resolution of most linear systems emanating from the application of the implicit, Crank-Nicolson, and implicit-explicit methods \cite{Leveque2007}.

Boundary conditions (BCs) in FDM are applied to the grid nodes lying on the boundary of $\Omega$ (i.e., $\partial\Omega$). The usual approach is to fix the value of the PDE solution on them (Dirichlet BCs) or to approximate a differential boundary condition with an FDM scheme (Neumman and Robin BCs) \cite{Leveque2007}. However, organ borders have complex geometries that rarely coincide with the FDM cartesian grid. This leads to the partition of $\Omega$ into the physical domain, corresponding to the tumor-harboring organ, and a fictitious domain, as shown in Fig.~\ref{fig:fdm}. FDM codes label grid nodes as physical or fictitious and only solve the PDE on the former. Additionally, FDM codes need to identify the physical nodes closer to organ borders to apply BCs.

FDM may be appealing for image-based cancer models because of its simplicity, rapid implementation, and that the cartesian grid can naturally fit the voxel datasets obtained with the imaging technologies described in Section~\ref{sec:img}, as shown in Fig.~\ref{fig:fdm}.
However, the FDM neglects the approximation of the organ geometry and simply relies on placing sufficient grid nodes to capture the organ's border. This impedes an accurate implementation of BCs and may also compromise the resolution of geometry-sensitive problems (e.g., mechanics). FEA and IGA  overcome these limitations, also providing superior numerical results that are supported by a strong and rigorously demonstrated mathematical basis \cite{Hughes2000,Cottrell2009}.

\subsection{Finite Element Analysis and Isogeometric Analysis}\label{sec:feaiga}

\subsubsection{General framework}

The central constituents of FEA and IGA are (i) the weak or variational formulation of the strong form of the model, and (ii) a robust approximation of this variational formulation using finite-dimensional function spaces with powerful approximation properties \cite{Hughes2000,Cottrell2009}. To define and illustrate these ideas, let us start by considering the stationary heat equation over a certain physical domain $\Omega$:
\begin{equation}\label{eq:sheat}
\nabla\cdot\left(\kappa\nabla u(\vec{x})\right) + f(\vec{x}) = 0,
\end{equation}
where $\kappa$ is the constant heat conductivity, $u(\vec{x})$ is the spatial map of temperatures over $\Omega$, and $f(\vec{x})$ is a heat source. We further consider homogeneous Dirichlet BCs (i.e., $u(\vec{x})=0$ on $\partial\Omega$), which together with Eq.~\eqref{eq:sheat} constitute the \emph{strong form} of the problem. To derive the \emph{weak form} or variational formulation of this PDE model, we define the \emph{trial function space} $\mathcal{U}$, where the PDE solution resides, and the \emph{weighting function space} $\mathcal{V}$. To this end, we choose $\mathcal{U},\mathcal{V} \subset \mathcal{H}^1$, which is the Sobolev space of square-integrable functions with square-integrable first derivatives. Standard FEA and IGA follow a Bubnov-Galerkin approach. For our heat problem, this translates in $\mathcal{U}=\mathcal{V}$ with functions $u\in\mathcal{U}$ and $w\in\mathcal{V}$ verifying $u(\vec{x})=w(\vec{x})=0$ on $\partial\Omega$. The interested reader is referred to \cite{Hughes2000,Cottrell2009} for a rigorous construction of $\mathcal{U}$ and $\mathcal{V}$. We obtain the weak form of our heat problem as follows:  we multiply all terms in Eq.~\eqref{eq:sheat} by an arbitrary $w(\vec{x})\in\mathcal{V}$, integrate the PDE in space over $\Omega$, and integrate the diffusive term by parts using the divergence theorem recalling that $w(\vec{x})=0$ on $\partial\Omega$, which cancels the boundary integral. As a result, the weak form is
\begin{equation}\label{eq:wheat}
\int_\Omega \nabla w(\vec{x}) \cdot\left(\kappa\nabla u(\vec{x})\right) d\vec{x} - \int_\Omega w(\vec{x})f(\vec{x})d\vec{x} = 0,
\end{equation}
\noindent which accounts for both the PDE and the BCs \cite{Hughes2000,Cottrell2009}. Let us define the finite-dimensional subspaces $\mathcal{U}^h\subset\mathcal{U}$ and $\mathcal{V}^h\subset\mathcal{V}$ to approximate the infinite-dimensional spaces $\mathcal{U}$ and $\mathcal{V}$, respectively. We choose a set of basis functions $\{B_A(\vec{x})\}_{A=1,\ldots,n_p}$ spanning $\mathcal{U}^h$ and $\mathcal{V}^h$, where $n_p=\mbox{dim}(\mathcal{U}^h)=\mbox{dim}(\mathcal{V}^h)$. This enables us to \emph{discretize} the weak form in space.
Now, our aim is to find $u^h(\vec{x})=\sum_{B=1}^{n_p}u_B B_B(\vec{x})$ in $\mathcal{U}^h$ satisfying Eq.~\eqref{eq:wheat} for any $w^h(\vec{x})=\sum_{A=1}^{n_p}w_AB_A(\vec{x})$ in $\mathcal{V}^h=\mathcal{U}^h$. In these expressions, the coefficients $u_B$ and $w_A$ are real constants. Using the definition of $w^h$ and recalling that all $w_A$ are arbitrary, we can simplify Eq.~\eqref{eq:wheat} to the Galerkin form
\begin{equation}\label{eq:gheat}
\int_\Omega \nabla B_A(\vec{x}) \cdot\left(\kappa\nabla u^h(\vec{x})\right) d\vec{x} - \int_\Omega  B_A(\vec{x})f^h(\vec{x})d\vec{x} = 0,
\end{equation}
for all $A=1,\ldots,n_p$ and where $f^h(\vec{x})=\sum_{A=1}^{n_p}f_A B_A(\vec{x})$. Then, by  introducing $u^h(\vec{x})=\sum_{B=1}^{n_p}u_B B_B(\vec{x})$ in Eq.~\eqref{eq:gheat} and rearranging terms, we obtain
\begin{equation}\label{eq:ssheat}
\sum_{B=1}^{n_p} u_B \int_\Omega \nabla B_A(\vec{x}) \cdot\left(\kappa\nabla B_B(\vec{x})\right) d\vec{x} = \int_\Omega  B_A(\vec{x})f^h(\vec{x})d\vec{x} ,
\end{equation}
\noindent for all $A=1,\ldots,n_p$. Eq.~\eqref{eq:ssheat} corresponds to a linear algebraic system $\vec{K}\vec{U}=\vec{F}$, where $\vec{F}=\{F_A\}$ and  $\vec{K}=\{K_{AB}\}$ are given by
\begin{equation}\label{eq:mheat}
 F_A = \int_\Omega  B_A(\vec{x})f^h(\vec{x})d\vec{x} \quad\mbox{and} \quad
 K_{AB} = \int_\Omega \nabla B_A(\vec{x}) \cdot\left(\kappa\nabla B_B(\vec{x})\right) d\vec{x}.
\end{equation}
\noindent The solution $\vec{U}=\{u_B\}$ provides the coefficients to determine the FEA or IGA approximation $u^h(\vec{x})=\sum_{B=1}^{n_p}u_B B_B(\vec{x})$ to our original model in Eq.\eqref{eq:sheat}.  In this process, the construction of the finite spaces $\mathcal{U}^h$ and $\mathcal{V}^h$ along with the basis $\{B_A(\vec{x})\}_{A=1,\ldots,n_p}$ are key steps that ultimately control the convergence and accuracy of the numerical scheme, and that exhibit methodological differences between FEA and IGA.

\begin{figure}[!t]
\includegraphics[width=\linewidth]{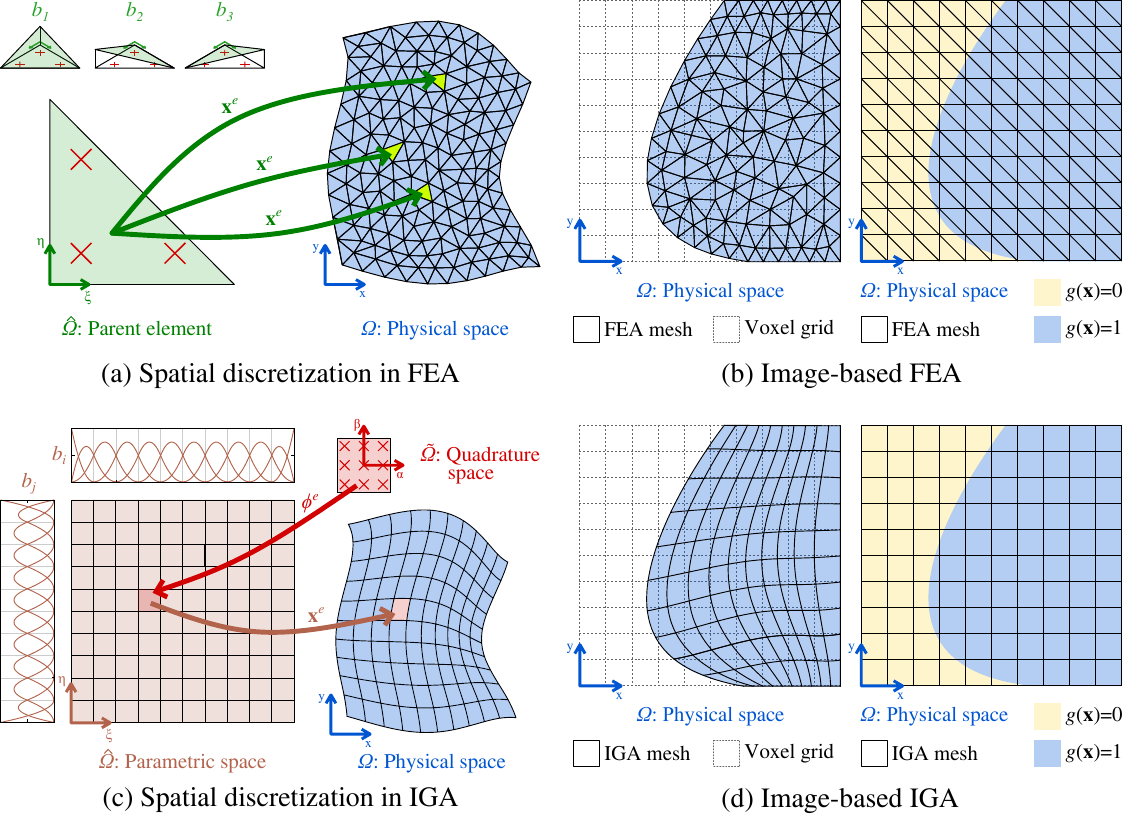}
\caption{Finite Element Analysis (FEA) and Isogeometric Analysis (IGA). 
(a) In FEA, we approximate the physical domain by repeatedly mapping a common parent element $\hat{\Omega}$ over $\Omega$ using a geometric map $\vec{x}^e$ to generate each element in the mesh. The parent element supports a local function basis $\{b_1,b_2,b_3\}$, which, once mapped to $\Omega$ from each element, contributes to the definition of a global  function basis. This is used to approximate the model equations in variational form and is usually integrated using quadrature rules also defined on the parent element (red crosses).
(b) Unstructured FEA meshes can be built to match the segmentation of an organ extracted from medical images (left). Alternatively, immersed-boundary approaches define a FEA mesh matching the voxel grid and a map $g(\mathbf{x})$ to identify the physical domain in which the model will be solved (right).
(c) In IGA, the physical domain is approximated with a topologically equivalent parametric space $\hat{\Omega}$, that is globally mapped onto $\Omega$. The parametric space results from the tensor product of univariate piecewise spline basis $\{b_i\}_{i=1,\ldots,n_i}$ and $\{b_j\}_{j=1,\ldots,n_j}$. The resulting multivariate spline basis is used to approximate the model equations in variational form. These can be integrated using quadrature rules (red crosses) defined over the quadrature space $\tilde{\Omega}$, which is mapped to $\Omega$ \textit{via} composition of the geometric maps $\phi^e$ and $\vec{x}^e$ for each element.
(d) IGA meshes can be built to match the segmentation of an organ extracted form medical images (left). Alternatively, immersed-boundary strategies define an IGA mesh aligning with the voxel grid and a map $g(\mathbf{x})$ to identify the physical domain in which the model will be solved (right).}
\label{fig:femiga}
\end{figure}

Standard FEA  uses piecewise Lagrangian polynomial bases to approximate $u^h(\vec{x})$ \cite{Hughes2000}. 
The piecewise architecture of FEA bases enables the partition of $\Omega$ in a mesh of $n_e$ subregions termed \emph{elements}, as shown in Fig.~\ref{fig:femiga}. 
FEA bases are also \emph{isoparametric} \cite{Hughes2000}, which is a crucial property enabling the use of the same basis functions to describe the geometry $\Omega$ of our problem  by means of a function $G(\vec{x})=\sum_{A=1}^{n_p} \vec{x}_A B_A(\vec{x})$, where $\vec{x}_A$ are the physical coordinates of a known set of points over the elements termed \emph{global nodes}.

FEA  bases $\{B_A(\vec{x})\}_{A=1,...,n_p}$ are built from a canonical \emph{local basis} defined on a \emph{parent element} $\hat{\Omega}$, which is common for all the elements in the mesh. 
The local basis  $\{b_a(\bm{\xi})\}_{a=1,...,n_b}$ is composed of $n_b$ functions constructed on the local coordinate system $\bm{\xi}$ of the parent element (see Fig.~\ref{fig:femiga}). 
Each local basis function $b_a$ is associated to a unique point in the parent element termed \emph{local node}, with local coordinates $\bm{\xi}_a$ (see Fig.~\ref{fig:femiga}). 
For each element $e$ in the mesh, we can build a \emph{geometric mapping} from the parent element given by $\vec{x}^e(\bm{\xi})=\sum_{a=1}^{n_b}\vec{x}^e_a b^e_a(\bm{\xi})$, where $\vec{x}^e_a$ are the physical coordinates of the local nodes of element $e$ in $\Omega$. Hence, we can repeatedly map the parent element and the local basis to each of the elements in the mesh (see Fig.~\ref{fig:femiga}), thereby obtaining the definition of the local basis over each element $e$, i.e.,  $\{b^e_a(\vec{x}^e)\}_{a=1,\ldots,n_b}$.
In this process, we also build a \emph{connectivity array} of the form $A=c(e,a)$ to identify the global nodes $\vec{x}_A$ and global basis functions $B_A(\vec{x})$ associated to each element's local nodes $\vec{x}^e_a$ and local functions $b^e_a(\vec{x}^e)$  \cite{Hughes2000}. Note that the combination of all element geometric mappings ultimately renders the geometric function $G(\vec{x})$ representing all $\Omega$.

The geometric map $\vec{x}^e$ is \emph{invertible}, such that we can use its inverse to map each element $e$ in the mesh back to the parent element. 
This \emph{pull-back} enables us to integrate any basis function over the common parent element and always use the same quadrature rule. 
This is a key idea to efficiently calculate vector $\vec{F}$ and matrix $\vec{K}$, using processes called \emph{formation} and \emph{assembly}. 
This strategy consists of looping over the elements of the mesh, such that for each element $e$ we (i) use the connectivity array to identify the local nodes and basis functions, (ii) pull them back to the parent element, (iii) calculate the integrals participating in Eq.~\eqref{eq:mheat} using Gaussian quadrature, and (iv) assemble the resulting local values $f_a^e$ and $K_{ab}^e$ by summing them into their corresponding global counterparts $F_A$ and $K_{AB}$ as indicated by the connectivity array. Note that $F_A$ and $K_{AB}$ will receive a contribution from each of the elements sharing node $A$.
In step (ii) we can also pull back spatially varying functions over each element, such as $f(\vec{x})$ or even $u^h(\vec{x})$ (e.g., in nonlinear problems).
Thus, the processes of formation and assembly capitalize on the piecewise definition of FEA bases over the elements to efficiently calculate the integrals in Eq.~\eqref{eq:mheat} specifically wherever the basis functions are defined, instead of performing an inefficient integration over the whole physical domain $\Omega$\cite{Hughes2000}. 

IGA is considered a generalization of FEA because it relies on the same core ideas. However, IGA employs more sophisticated polynomial functions coming from computer graphics because its root idea is to use the functions \emph{exactly} describing a computer-generated geometric model of $\Omega$ (e.g., an engineering design, an organ segmentation) to numerically solve the PDE problem posed on such geometry \cite{Cottrell2009}. Conversely, in FEA we first choose the basis to approximate the solution $u^h(\vec{x})$ and then we use it to describe  $\Omega$, which usually results in an approximation of $\Omega$ as well. Thus, IGA bases are \emph{geometrically-exact} and isoparametric. IGA bases also show higher global continuity, which ultimately yields superior accuracy \cite{Cottrell2009}.

Non-Uniform Rational B-splines (NURBS) define the most usual function space in IGA \cite{Cottrell2009,Farin1999}.
Univariate NURBS bases are globally defined by a \emph{knot vector}, which is a set of non-decreasing coordinates termed \emph{knots} enabling the definition of all NURBS basis functions over a segment \emph{parametric space} $\hat{\Omega}$ (see Fig.~\ref{fig:femiga}). Multivariate NURBS bases are defined by the tensor product of univariate NURBS bases. Likewise, the tensor product of the corresponding knot vectors results in the definition of the complete IGA mesh formed by $n_e$ elements in a multivariate \emph{parametric space} $\hat{\Omega}$, which is topologically equivalent to the physical space $\Omega$. The isogeometric elements are formed by the knot lines in each parametric direction (see Fig.~\ref{fig:femiga}). 
Each of the resulting $n_p$ basis functions $B_A(\bm{\xi})$ defined in $\hat{\Omega}$ is associated to a \emph{control point}, with parametric coordinates $\bm{\xi}_A$ and physical coordinates $\vec{x}_A$. 
This results in the definition of an invertible \emph{global geometric mapping}  $\vec{x}^g(\bm{\xi})=\sum_{A=1}^{n_p} \vec{x}_A B_A(\bm{\xi})$, bringing the whole IGA mesh from the parametric space $\hat{\Omega}$ into the physical space $\Omega$ and providing an explicit definition of the problem geometry (i.e., $G(\vec{x})$). 
Contrary to global nodes in FEA, the control points in IGA do not necessarily align with the mesh and may even be placed out of $\Omega$ \cite{Farin1999,Cottrell2009}.
The restriction of $\vec{x}^g$ to each of the elements also enables the construction of the invertible \emph{element geometric mapping} $\vec{x}^e$, which relies on the identification of the $n_b$ local basis functions and associated local control points defined over the element $e$ by means of a \emph{connectivity array} as in FEA. Additionally, the connectivity array in IGA further accounts for the univariate basis functions that gave rise to the multivariate basis functions \cite{Cottrell2009}.

As the knot vectors are arbitrary, the elements in parametric space may have varying sizes. Thus, we further define a unique \emph{quadrature space} $\tilde{\Omega}$, which is common to all elements, and an ancillary invertible element mapping $\phi^e(\tilde{\bm{\xi}})$ from $\tilde{\Omega}$ to each element in $\hat{\Omega}$. During the process of \emph{formation} and \emph{assembly}, we combine $\vec{x}^e$ and $\phi^e$ to perform the \emph{pull-back} from $\Omega$ to $\tilde{\Omega}$ and calculate the integrals participating in matrices and vectors of the final system (see Eq.~\eqref{eq:mheat}). The rest of the assembly steps in IGA are essentially the same as in FEA (see \cite{Cottrell2009}).

\subsubsection{FEA and IGA for image-based cancer models}

Let us now consider the reaction-diffusion cancer growth model in Eq.~\eqref{eq:rd}. Standard FEA and IGA approaches are \emph{boundary-fitted}, i.e., the physical space $\Omega$ represents the patient's organ (see Fig.~\ref{fig:femiga}). Here, we will derive the weak form of Eq.~\eqref{eq:rd} with the usual no-flux BC $\nabla N\cdot\vec{n}=0$. We choose $\mathcal{U},\mathcal{V} \subset \mathcal{H}^1$ and, using a Bubnov-Galerkin approach, we obtain $\mathcal{U}=\mathcal{V}$ \cite{Hughes2000,Cottrell2009}. 
We define $\mathcal{U}^h\subset\mathcal{U}$ and $\mathcal{V}^h\subset\mathcal{V}$ and we choose a basis $\{B_A(\vec{x})\}_{A=1,\ldots,n_p}$ to span $\mathcal{U}^h$ and $\mathcal{V}^h$. Standard FEA and IGA approximate time-dependent functions $f(\vec{x},t)$ as $f^h(\vec{x},t)=\sum_{A=1}^{n_p}f_A(t)B_A(\vec{x})$. Then, the Galerkin discretization of the reaction-diffusion model's weak form is
\begin{equation}\label{eq:galerkin}
\int_{\Omega} B_A\frac{\partial N^h}{\partial t} d\Omega + \int_{\Omega} \nabla B_A \cdot(D\nabla N^h) d\Omega - \int_{\Omega} B_A k^h N^h \left(1-\frac{N^h}{\theta} \right) d\Omega = 0,
\end{equation}
for all $A=1,\ldots,n_p$ and where $k^h(\vec{x})=\sum_{A=1}^{n_p}k_A B_A(\vec{x})$. To discretize and integrate in time, we propose the \emph{generalized-$\alpha$ method} \cite{Chung1993,Jansen2000,Cottrell2009}. This approach relies on a partition of $\left[0,T\right]$ in subintervals $[t_n, t_{n+1}]$ with an arbitrary time step $\Delta t_n$,  the discretization of $\partial N^h(\vec{x},t)/\partial t = \sum_{A=1}^{n_p}\dot{N}_A(t)B_A(\vec{x})$, and the definition of vectors $\vec{N}^n=\vec{N}(t_n)=\{ N_A(t_n)\}$ and  $\dot{\vec{N}}^n=\dot{\vec{N}}(t_n)=\{ \dot{N}_A(t_n)\}$. The generalized-$\alpha$ method is a predictor-multicorrector algorithm that provides $\vec{N}^{n+1}$ and $\dot{\vec{N}}^{n+1}$ given $\vec{N}^n$ and $\dot{\vec{N}}^n$. This method can be proven to be second-order accurate and $A$-stable by adequately choosing its defining parameters (see \cite{Cottrell2009,Colli2020}). The resulting system of nonlinear algebraic equations can be solved iteratively with Newton's method (see \cite{Cottrell2009,Colli2020}).

Finally, there are multiple approaches to generate \emph{boundary-fitted} meshes in FEA and IGA \cite{Zhang2018}. \emph{Unstructured meshes} are the usual strategy in FEA and consist of populating a geometric model of the patient's tumor-hosting organ generated from its segmentation with tetrahedral or hexahedral elements \cite{Agosti2018,Weis2013,lima17}. In IGA, a common approach is \emph{parametric mapping}, whereby a known geometric model that is topologically equivalent to the organ's geometry is deformed to match the organ's segmentation \cite{Lorenzo2016,Lorenzo2019}. However, image-based cancer models may be more amenable to \emph{immersed-boundary} methods \cite{Mittal2005,Burman2015,Schillinger2015,Parvizian2007,Duester2008}. These strategies rely on constructing the FEA/IGA mesh to align with the voxel grid and  defining a function $g(\vec{x})$ accurately representing the organ's boundary. Hence, the function $g(\vec{x})$ enables the definition of the physical space $\Omega$ and a fictitious space over the rest of the mesh (see Fig.~\ref{fig:femiga}). The model PDEs are solved over the whole mesh but their parameters are weighted with $g(\vec{x})$, such that they take their usual value in the physical domain and a negligible value in the fictitious domain. Immersed-boundary methods may also rely on local refinement to improve the discretization of $g(\vec{x})$  \cite{Schillinger2015}.

\section{Calibrating image-based mathematical oncology models}\label{sec:invprob}

\begin{figure}[t]
    \centering
    \includegraphics[width=\textwidth]{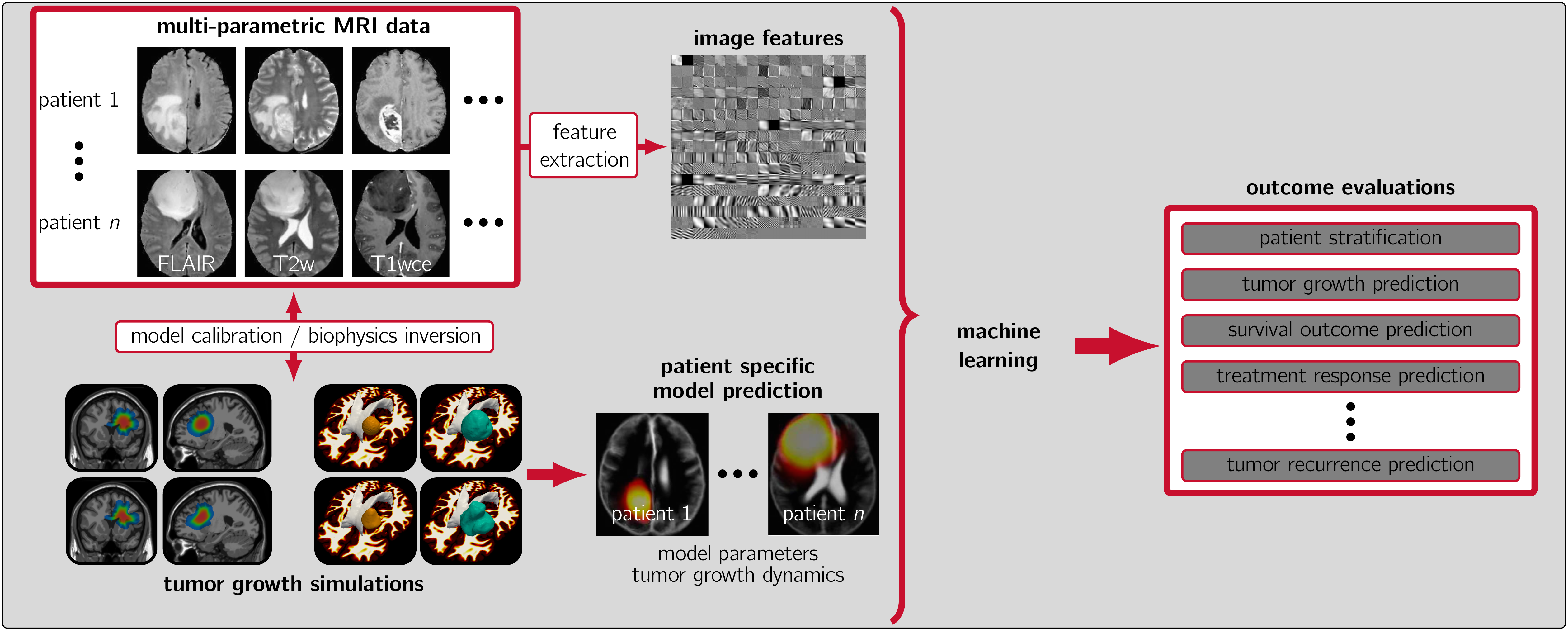}\\ (a) Imaging workflow in brain tumor radiomics \vspace{2em}\\
    \includegraphics[width=\textwidth]{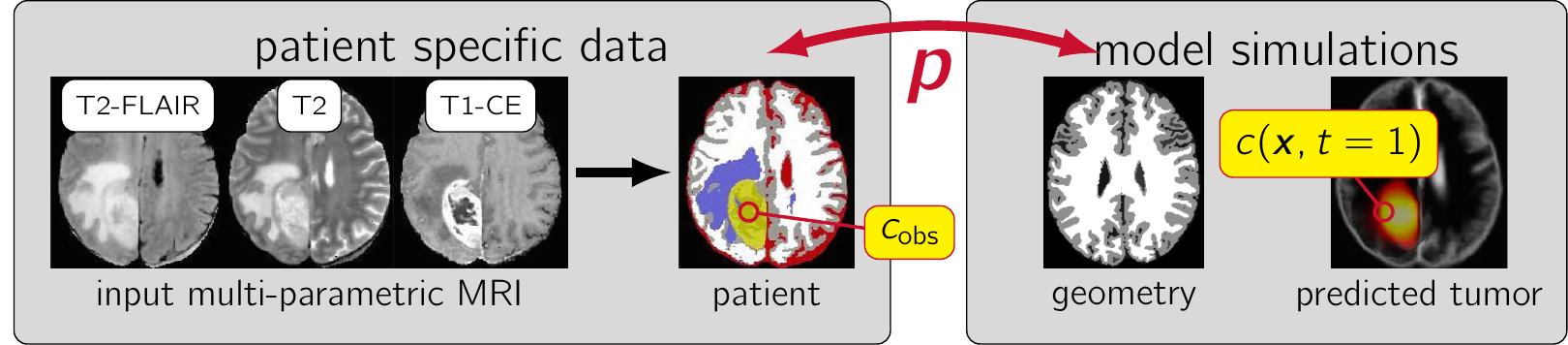}\\ (b) Inverse problem \\
    \caption{Workflow for an image-driven biophysical inverse problem. (a) The goal is to extract clinically relevant biomarkers from input multi-parametric MRI data that can predict and guide intervention so as to improve clinical outcome. Purely imaging-based approaches (feature extraction) can be integrated with biophysical priors (biophysical model inversion) to develop tools that can assist in treatment and prognosis.  (b) The input data for patients with brain tumors (multi-parametric MRI scans) is at a single time point (treatment typically follows immediately after diagnosis) and is translated into model observables for tumor concentration $c_\text{obs}$.        
         The inverse problem seeks to estimate unknown biophysical parameters $\vec{p}$ of a given mathematical model. The basic idea is to perform a number of model simulations with different parameters so that the model-predicted tumor matches the observed data. The panels in this figure are adapted from~\cite{Mang:2020}.}\label{fig:model-calibration}
\end{figure}

Mathematical models of cancer define a \emph{forward problem}, whose solution provides \emph{state variables} (e.g., tumor cell density). In general, these models are parameterized by \emph{unknown} biophysical parameters (and possibly initial conditions) that typically manifest substantial variability across subjects~\cite{Swanson:2020a,Lipkova:2019a,Scheufele:2020}. 
The estimation of these unknown variables (also called \emph{inversion variables}) should be patient-specific and can be mathematically posed as an \emph{inverse problem}, which aims at optimizing an objective function constrained by the model. 
Since image-based cancer models are usually represented by PDEs, the resulting inverse problem is formally a PDE-constrained optimization problem.
In this section, we outline the general formulation of the inverse parameter estimation problem and discuss the standard methods to compute its solution, keeping in mind the ultimate goal of patient-specific tumor characterization and model prediction. Fig.~\ref{fig:model-calibration} illustrates the typical image-based inverse problem workflow in the context of brain tumors.

\subsection{Inverse problems for oncology models}
Let $\mathcal{F}(\vec{p},c) = 0$ denote a cancer model consisting of a PDE system, where $\vec{p}$ is a vector of unknown parameters and $c(\vec{x},t)$ represents the state variables.
The inverse problem seeks to estimate $\vec{p}$ such that the model state variables $c(\vec{x},t)$ match given observational patient data (see Fig.~\ref{fig:model-calibration}). 
In oncology, the input data is usually a series of medical images (e.g., MRI, PET) at specific time instances $\{t_j\}_{j = 1, \cdots, n_t}$ within a finite time horizon $(0,T]$.
These define our observations  $\hat{c}_j(\vec{x})$ of the state variables in the forward model $c(\vec{x},t)$, e.g., tumor cell density (see Section~\ref{sec:dwi}).
Then, the inverse problem can be mathematically formulated as
\begin{equation}  \label{eq:pdeopt}
\min_{\vec{p}}
\mathcal{J}(\vec{p},c) = \frac{1}{2} \sum_{j=1}^{n_t} \int_{\Omega}\!\! \big(c(\vec{x},t_j) - \hat{c}_j(\vec{x})\big)^2 d \vec{x}  + \mathcal{R}(\vec{p}) \;\; \text{s.t.} \;\; \mathcal{F}(\vec{p},c) = 0,
\end{equation}
where $\Omega$ is the spatial domain representing the patient's tumor-harboring organ. The objective function $\mathcal{J}(\vec{p},c)$ minimizes the mismatch between the predicted tumor and observed data at times $\{t_j\}_{j = 1, \cdots, n_t}$ using an $L^2$ distance measure. Additionally, it balances this data fidelity/mismatch term with a regularization operator $\mathcal{R}(\vec{p})$. 

Designing algorithms for the efficient and effective numerical solution of such PDE-constrained optimization problems is a challenging task~\cite{Leugering:2014a,Akcelik:2006a}. While derivative-free optimization strategies are popular due to ease of implementation, they are typically associated with slow convergence and can become prohibitively expensive (especially if the dimensionality of $\vec{p}$ is large). Hence, optimization algorithms that utilize gradient information are preferable. In addition to improving convergence and computation time, these methods can reveal important characteristics of the objective function landscape, which can be exploited to design better algorithms and help us understand the sensitivities and ill-posedness inherent to PDE-constrained optimization problems.
There exist multiple options for evaluating the gradient (and higher order derivatives) of the objective function, such as automatic differentiation, numerical approximation through finite differences, and adjoint-based methods.
For oncology models, several groups have employed adjoints for inversion~\cite{Hogea:2008b,Gholami:2016a,Subramanian20IP,Colin:2014a,Feng:2018a,Knopoff:2013a}. Some efforts also employ Hessian information to accelerate convergence~\cite{Gholami:2016a,Scheufele:2018a}. Other strategies such as derivative-free optimization~\cite{Chen:2012b,Wong:2017a,Mi:2014a,Konukoglu:2010b} or finite difference approximations~\cite{Hormuth:2015a} have also been considered in literature, but are usually less effective than adjoint-based methods. For large-scale 3D inversion, parallel (distributed memory) algorithms have been considered in~\cite{Gholami:2016a,Scheufele:2018a,Subramanian20IP}. 

While the deterministic approaches just described are successful in estimating the optimal parameters to the minimization problem in Eq.~\eqref{eq:pdeopt}, their utility could be limited due to uncertainties arising from modeling errors and noise in measurements/data. A probabilistic (Bayesian) formulation can mitigate this drawback by characterizing our confidence in the inversion variables $\vec{p}$ using probability density functions. This approach will be described in Section~\ref{sec:probcal}, as part of a comprehensive Bayesian framework for model selection. In the following section, we briefly describe adjoint-based inversion methods.

\subsection{Adjoint methods for inverse problems}
The standard technique for solving the inverse problem posed in Eq.~\eqref{eq:pdeopt} is to introduce Lagrange multipliers $\lambda(\vec{x},t)$, which are termed \emph{adjoint variables} or simply \emph{adjoints}, and construct the Lagrangian functional as
\begin{equation}\label{eq:lagrangian}
\mathcal{L}(c, \vec{p}, \lambda) = \mathcal{J}(\vec{p},c) + 	\langle \mathcal{F}(\vec{p},c), \lambda \rangle,
\end{equation}
where $\langle . \rangle$ denotes an appropriate inner product (typically $L^2$) and $\mathcal{J}(\vec{p},c)$ denotes the objective function defined in Eq.~\eqref{eq:pdeopt}. By requiring stationarity with respect to the state, adjoint, and inversion variables, we arrive at the \emph{first order optimality conditions} by taking the following variations:
\begin{subequations}\label{eq:opt}
\begin{align}
         \delta_\lambda \mathcal{L} &= 0 \quad \quad\quad \text{\textit{(forward equations)}},\label{eq:forward}\\
    \delta_c \mathcal{L} &= 0 \quad \quad\quad \text{\textit{(adjoint equations)}},\label{eq:adjoint}\\
      \delta_{\vec{p}} \mathcal{L} &= 0 \quad \quad\quad \text{\textit{(inversion equations)}},\label{eq:inversion}
\end{align}
\end{subequations}
where $\delta_z  \mathcal{L}$ denotes the variation of $ \mathcal{L}$ with respect to $z$. The forward equations are simply the tumor growth PDE model. The adjoint equations are linear PDEs in the adjoint variables backward in time. The inversion equations denote the PDE-constrained gradient of $\mathcal{J}(\vec{p},c)$, which is set to zero at the local minimum. 

For example, consider the reaction-diffusion cancer model in Eq.~\eqref{eq:rd} with the usual no-flux BC $\nabla N\cdot\vec{n}=0$ and two tumor cell density observations $\hat{N}_0(\vec{x})$ and $\hat{N}_1(\vec{x})$ derived from DW-MRI at $t=t_0=0$ and $t=t_1$  (see Section~\ref{sec:dwi}). We assume that $\hat{N}_0(\vec{x})$ are known initial conditions for  Eq.~\eqref{eq:rd}, and use $\hat{N}_1(\vec{x})$ to calibrate $\vec{p}=\{D,k(\vec{x})\}$, where $D$ is a scalar constant and $k(\vec{x})$ is a spatial field, representing the tumor cell diffusion and proliferation rate, respectively  (see Section~\ref{sec:dwi}). Then, we define the objective function $\mathcal{J}(\vec{p},N)$ and the Lagrangian functional $\mathcal{L}(N, \vec{p}, \lambda)$ as
\begin{align}\label{eq:glL}
\mathcal{J}(\vec{p},N) &= \frac{1}{2} \int_{\Omega}\!\! \big(N(\vec{x},t_1) - \hat{N}_1(\vec{x})\big)^2 d \vec{x}  + \frac{a}{2}\left(D^2 + \int_\Omega\!\! k^2(\vec{x})d\vec{x}\right), \\
\mathcal{L}(N, \vec{p}, \lambda) &= \mathcal{J}(\vec{p},N) + \int_0^{t_1}\int_\Omega\!\!\lambda(\vec{x},t)\left( \frac{\partial N(\vec{x},t)}{\partial t}  - \nabla\cdot\left(D \nabla N(\vec{x},t)\right)  \right. \notag \\ & \left. - k(\vec{x})N(\vec{x},t)\left(1-\frac{N(\vec{x},t)}{\theta}\right)  \right)d\vec{x}dt,
\end{align}
where $a$ is a regularization parameter. Following Eq.~\eqref{eq:adjoint}, the adjoint equation is
\begin{equation}
-\frac{\partial\lambda(\vec{x},t)}{\partial t} = \nabla\cdot\left(D \nabla \lambda(\vec{x},t)\right) + k(\vec{x})\lambda(\vec{x},t)\left(1-\frac{2N(\vec{x},t)}{\theta}\right),
\end{equation}
which is a linear backward problem in time subject to the BC $\nabla \lambda\cdot\vec{n}=0$ on $\partial\Omega$ and the terminal condition $\lambda(\vec{x},t_1)=\hat{N}_1(\vec{x})-N(\vec{x},t_1) $.  Following Eq.~\eqref{eq:inversion}, the inversion equations are given by
\begin{align}
  aD + \int_0^{t_1}\int_\Omega\!\! \nabla\lambda(\vec{x},t) \cdot \nabla N(\vec{x},t) d\vec{x}dt &= 0, \\
  ak(\vec{x}) -  \int_0^{t_1}\!\! \lambda(\vec{x},t) N(\vec{x},t)\left(1-\frac{N(\vec{x},t)}{\theta}\right)dt &= 0.
\end{align}
 
In general, Eq.~\eqref{eq:opt} represents a large, non-linear, coupled system of PDEs, which can be significantly challenging to solve simultaneously. Instead, a standard approach involves a \emph{reduced space} algorithm, which is an iterative strategy that reduces Eq.~\eqref{eq:opt} to a system involving only the inversion variables. In each iteration, we use the current approximation to $\vec{p}$ to solve the state and adjoint equations to respectively get the current approximation to the state and adjoint variables. Then, we update our inversion variables using the inversion equations. This process is repeated until convergence, which is typically set by a user-defined threshold on the parameter update (i.e., the parameter gradient). In contrast to \emph{full space} methods, reduced space methods present more tractable systems that can exploit existing PDE solvers for the state and adjoint equations and are better conditioned. We refer the reader to~\cite{Akcelik:2006a} for more details on adjoint methods in PDE-constrained optimization.

\section{Model selection and identification of relevant parameters}

Given the vast array of cancer growth models in the literature, it is not trivial to choose which is the best to represent the available data and to predict key quantities of interest (e.g., the tumor size, treatment efficacy, or percentage of necrosis) for a certain tumor type.
In this section, we describe the Occam Plausibility Algorithm (OPAL), which has been proposed in \cite{Farrell2015} as an adaptive process for model selection and validation in the presence of uncertainties. The strategy relies on three key steps: sensitivity analysis to identify the relevant model parameters, model calibration of the relevant parameters, and calculation of model selection criteria. The OPAL can be referred to as \emph{model agnostic} in that no single model is advocated; rather, the \emph{best} model is selected based upon our model selection criteria. 
Details regarding each step of the OPAL are given in the next subsections.

\subsection{Variance-based sensitivity analysis}\label{sec:sobol}

Sensitivity analysis quantifies how changes in parameter values affect the uncertainty in model output \cite{saltelli2019so}. We can distinguish between \emph{local} and \emph{global} methods. Local methods compute the variation of the model output changing one parameter at a time (i.e., first-order effects) usually \emph{via} derivation, but neglect the interactions between the parameters. 
In global methods, the contribution of each parameter along with its interactions with other parameters (i.e., higher-order effects) are taken into account, as all parameters are varied simultaneously over the entire parameter space. In this work, we present the variance-based global sensitivity analysis method, also known as the Sobol method \cite{sobol2001global,saltelli2008global}. 
Details regarding local and other global methods can be found in \cite{saltelli2008global}.

Let $\vec{M}(\bm{\theta})$ be a model parameterized by $k$ parameters $\bm{\theta}$, which belong to a parameter space $\bm{\Theta}\subset\mathbb{R}^k$. The computational cost of the sensitivity analysis of model $\vec{M}(\bm{\theta})$ depends on the number of parameters $k$ and the sample size $N$, with the total number of model evaluations given by $N_T=N(k+1)$. There are several approaches to estimate the \emph{total sensitivity index} for each parameter, which quantifies all effects of the parameter on the model output. Here, we present the strategy in  \cite{saltelli2010variance}, as it is known to demand a reduced sample size to converge.

First, we randomly generate two sampling matrices, $\vec{A}$ and $\vec{B}$, with size $N\times k$. Each row of these matrices represents a sampled value for the vector of parameter $\bm{\theta}$. Additionally, we create $k$ matrices $\vec{A}_{\vec{B}}^{(k)}$, where we copy the values from the matrix $\vec{A}$ and replace the values from column $k$ with the values from $\vec{B}$. For the case where $N=1$, these matrices are given as: $\vec{A}=[ \theta^a_{1,1} \quad \theta^a_{1,2} \; \ldots \; \theta^a_{1,k}]$, $\vec{B}=[ \theta^b_{1,1} \quad \theta^b_{1,2} \; \ldots \; \theta^b_{1,k}]$,  $\vec{A}_{\vec{B}}^{(1)}=[ \theta^b_{1,1} \quad \theta^a_{1,2} \; \ldots \; \theta^a_{1,k}]$, $\vec{A}_{\vec{B}}^{(2)}=[ \theta^a_{1,1} \quad \theta^b_{1,2} \; \ldots \; \theta^a_{1,k}]$, $\ldots$, and $\vec{A}_{\vec{B}}^{(k)}=[ \theta^a_{1,1} \quad \theta^a_{1,2} \; \ldots \; \theta^b_{1,k}]$.

Then, we run the forward model for each row in matrix $\vec{A}$ and all matrices $\vec{A}_{\vec{B}}^{(k)}$.
The outputs of the model are stored in corresponding solution vectors for each matrix; i.e., $\vec{Y}_A$, $\vec{Y}_{AB}^1$, \ldots, $\vec{Y}_{AB}^k$. Finally, we compute the total sensitivity index $S_{T_i}$ for each parameter, which can be approximated \cite{jansen1999analysis,saltelli2008global,saltelli2010variance} by
\begin{align}
S_{T_i}=\frac{1}{2N\mathrm{Var}(\vec{Y}_{A})}\sum_{j=1}^{N}\left(\left(\vec{Y}_{A}\right)_j-\left(\vec{Y}_{AB}^{(i)}\right)_j\right)^2,
\label{eq_tei}
\end{align}
where $\mathrm{Var}(\vec{Y}_{A})$ is the variance of vector $\vec{Y}_{A}$. According to \cite{saltelli2008global}, a parameter $i$ can be considered non-influential if $S_{T_i}=0$. In practice, we define a threshold $\epsilon$ and we identify a parameter $i$ as non-influential if $S_{T_i}<\epsilon$. The choice of $\epsilon$ is relative to other $S_{T_j}$ and problem-dependent.
If $S_{T_i}$ is sufficiently small, then the parameter does not affect the quantities of interest, and the complexity of the model can be reduced by removing or fixing the parameter to any value within the uncertainty range \cite{saltelli2008global}. 

\subsection{Model calibration}\label{sec:probcal}

To characterize uncertainties in the observable data and the stochastic behavior of tumor growth, we follow the Bayesian statistical calibration procedure. This method captures these uncertainties by delivering a probabilistic distribution of the model parameters, instead of a single value for each of them \cite{oden2016,lima17,Lima2016,oden2017df}.  The basic ideas behind the Bayesian parameter estimation involve the following steps:
\begin{enumerate}
\item Select the observational data $\vec{D}$ to be used (e.g., baseline and follow-up MRI).

\item Establish the prior distribution of the model parameters $\pi_{prior}(\bm{\theta})$. In the cases where we do not have knowledge regarding the distribution of the parameters, and we can only estimate the range of these parameters, the usual approach is to assume a uniform prior distribution.

\item Construct the likelihood function, which, given the values assigned to the parameters $\bm{\theta}$, yields the probability of $\vec{D}$ being observed \cite{oden2017df}. Assuming both the experimental error and the model inadequacy to be Gaussian, and the experimental data to be independent and identically distributed, the likelihood is given as
\begin{align}
\pi_{like}(\vec{D}|\bm{\theta})=\prod_{i=1}^{N_t}\frac{1}{\sqrt{2\pi \sigma^2}}\exp\left(-\frac{(D_i-Y_i(\bm{\theta}))^2}{2\sigma^2}\right),
\label{eq:like}
\end{align}
where $N_t$ is the number of data points, $\sigma$ is the standard deviation of the experimental error and model inadequacy, and $Y(\bm{\theta})$ is the model output.
\item Compute the posterior distribution of the parameters $\pi_{post}(\bm{\theta}|\vec{D})$ as
\begin{align}
\pi_{post}(\bm{\theta}|\vec{D})=\frac{\pi_{like}(\vec{D}|\bm{\theta})\pi_{prior}(\bm{\theta})}{\pi_{evid}(\vec{D})},
\label{eq:post}
\end{align}
where $\pi_{evid}(\vec{D})=\int_\Theta\left(\pi_{like}(\vec{D}|\bm{\theta})\pi_{prior}(\bm{\theta})\right)\,d\bm{\theta}$ is the model evidence. The resulting posterior distribution of the parameters allows the prediction of the quantities of interest taking into account the uncertainties in the parameters.
\end{enumerate}

The posterior probability density function is, in general, non-Gaussian. Sampling schemes such as Markov Chain Monte Carlo (MCMC) methods can be used to evaluate posterior expectations \cite{lima17,Lima2016}. These stochastic methods can incur in a large computational cost, so efficient sampling strategies exploiting the problem structure are actively being investigated \cite{prudencio12,Drzisga2017,Wang2018,Petra2014}.

\subsection{Model selection criteria}

Following the Bayesian framework used for calibration, we approach model selection by  computing the \emph{model plausibility} \cite{jeffreys98,chow81,beck10,prudencio12,oden2017df}.
Given a set of $m$ models $\vec{M}=\left\{M_i(\bm{\theta}_i)\right\}_{i=1}^m$, the Bayes' rule in Eq. (\ref{eq:post})  can be rewritten assuming that probabilities are conditional on the model $M_i$ and the set ${M}$:
\begin{align}
\pi_{post}(\bm{\theta}_i|\vec{D},M_i,\vec{M})=\frac{\pi_{like}(\vec{D}|\bm{\theta}_i,M_i,\vec{M})\pi_{prior}(\bm{\theta}_i|M_i,\vec{M})}{\pi_{evid}(\vec{D}|M_i,\vec{M})}.
\label{eq:pospm}
\end{align}
The evidence of each model can be viewed as a likelihood for a discrete Bayesian calculation, yielding a new posterior called the model plausibility, $\pi_{plaus}$, given as
\begin{align}
\pi_{plaus}(M_i|\vec{M},\vec{D})=\frac{\pi_{evid}(\vec{D}|,M_i,\vec{M})\pi_{prior}(M_i|\vec{M})}{\pi_{evid}(\vec{D}|\vec{M})},\qquad 1\leq i \leq m.
\label{eg:plaupm}
\end{align}
If we assume that all models are equally probable, $\pi_{prior}(M_i|\vec{M})=\frac{1}{m}$. The sum of all model plausibilities is equal to one (i.e., $\sum_{i=1}^m\pi_{plaus}(M_i|\vec{M},\vec{D})=1$). The model with highest plausibility is selected as the best model in  $\vec{M}$ to capture the data.

Another popular method of model selection is the Akaike Information Criterion (AIC) \cite{konishi08}. In this method, the likelihood of the maximum likelihood estimator, $\hat{\bm{\theta}}$, is penalized according to the number of model parameters, i.e., 
\begin{align}
AIC_i=-2\log \pi_{like}(\vec{D}|\hat{\bm{\theta}})+2k_i, \qquad 1\leq i \leq m,
\end{align}
where $k_i$ the number of parameters in model $M_i$. In this case, the model with the lowest AIC is the best model in $\vec{M}$. 

\subsection{The Occam Plausibility Algorithm }

\begin{figure}[t]
\centering
\includegraphics[width=1.0\hsize]{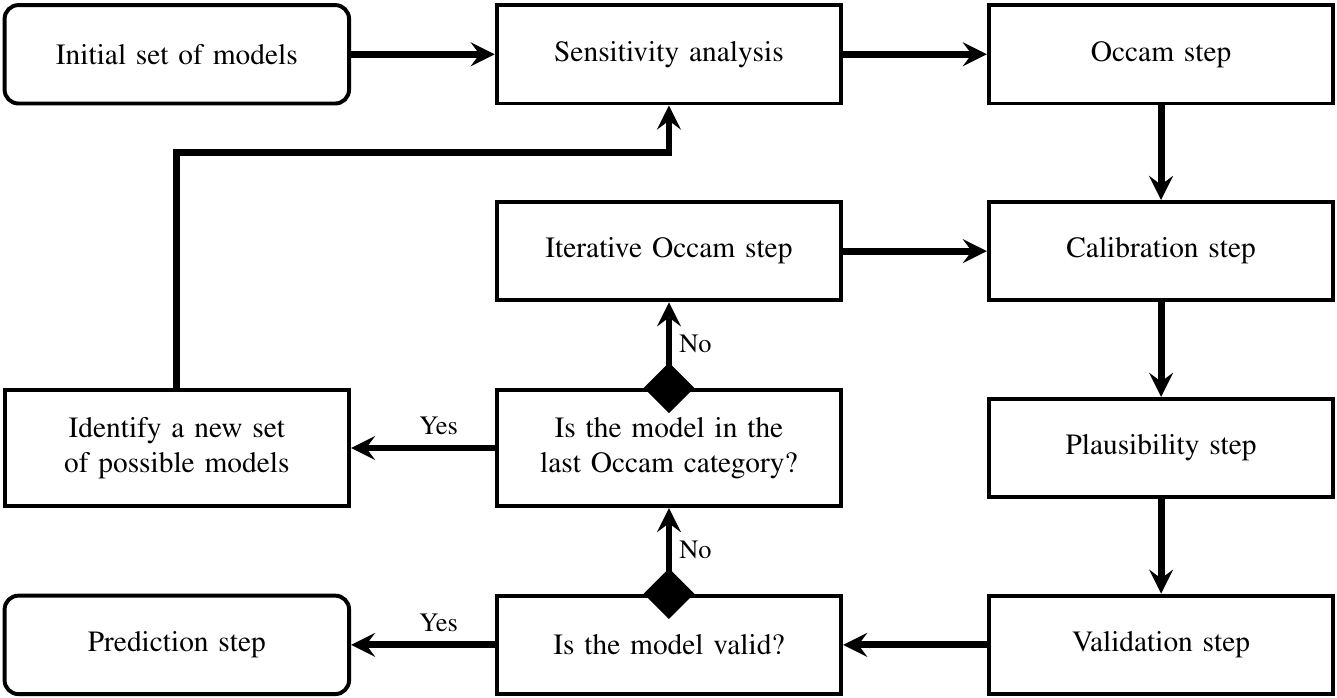}
\caption{The Occam Plausibility Algorithm (OPAL). This framework brings  together experimental data for model calibration and validation, sensitivity analysis, selection, and model prediction.}
\label{fig:opalf}
\end{figure}

OPAL has been proposed as a comprehensive framework that combines parameter sensitivity analysis, and model calibration, validation, selection, and prediction \cite{Farrell2015,oden2017df}. Fig. \ref{fig:opalf} shows a schematic representation of the main OPAL steps, which we outline in the following \cite{Farrell2015,oden2016}:
\begin{enumerate}
\item Identify a set $\vec{M}=\{M_1(\bm{\theta}_1),M_2(\bm{\theta}_2),\ldots,M_m(\bm{\theta}_m)\}$ of parametric models, each with parameters $\bm{\theta}_i$ belonging to an appropriate parameter space $\bm{\Theta}_i$, $1\leq i \leq m$.

\item Perform a sensitivity analysis to identify non-influential parameters. Based on this analysis, the values of these parameters are fixed to the mean value used in the sensitivity analysis. If there is a model in $\vec{M}$, whose only difference to other model is given by the non-influential parameters, this model can be eliminated, yielding a reduced set $\bar{\vec{M}}=\{\bar{M}_1(\bm{\theta}_1),\bar{M}_2(\bm{\theta}_2),\ldots,\bar{M}_l(\bm{\theta}_l)\}$ of models, with $l\leq m$.

\item Divide the models in $\bar{\vec{M}}$ into ``Occam Categories'' according to their complexity (e.g., number of parameters). These categories are sorted in ascending order according to their complexity.

\item Calibrate the models in Category 1 using the calibration data $\vec{D}^c$.

\item Select the best model to represent the data in this category (e.g., the model  with the highest plausibility).

\item Test the best model identified in the current category in a validation scenario, where  the posterior from the calibration step is used as a prior and the distribution of the parameters are updated against the validation data $\vec{D}^v$. If the model is able to represent the data within a preset tolerance, the model is considered ``valid''. If not, we return to step 3 and move to the next Occam category. If we are not able to find a valid model, we need to return to step 1 and include new models.

\item After finding the ``simplest'' valid model, solve the forward model in the prediction scenario and compute the quantities of interest.
\end{enumerate}

All of these steps are designed to consider uncertainties in the choice of model, the model parameters, the observational data, and the target quantities of interest. All uncertainties are generally characterized by probability densities.

\section{Towards the optimization of personalized treatment plans}\label{sec:optcon}

Several image-based mathematical models of cancer growth have shown promise in predicting  treatment outcomes in a patient-specific manner, as discussed in Sections~\ref{sec:rt} and \ref{sec:ct}. Those models could provide a means to determine optimal therapeutic regimens to treat a certain type of cancer \emph{in silico}, which could then be investigated within a clinical trial \emph{in vivo}. Hence, this computational approach seeks to help clinicians navigate the vast array of radiotherapy and drug combinations, dosing options, and treatment schedules and select optimal strategies, which are virtually impossible to assess in clinical trials. Ultimately, cancer models could also serve as a \emph{digital twin} for the patient's tumor, thereby enabling the pathological assessment, monitoring, and design of optimal therapeutic regimens for the individual patient \emph{in silico}.
In this section, we discuss the use of image-based predictive tumor growth models accounting for the therapeutic regimen and associated tumor response for the discovery of optimal therapeutic regimens and the design of patient-specific optimal treatment strategies.

\subsection{Potential to select treatment plans for individual patients}\label{sec:8.2}

Selecting a treatment regimen for a patient is a complex process. Oncologists use decision tree algorithms to select therapeutics for each patient considering, for example, tumor grade and cell markers \cite{Bevers2009}. However, the determination of the optimal dosing regimens for these therapies is vastly underinvestigated. This limitation follows from the impossibility to test all the potential dosing strategies within a clinical trial. Additionally, regimens may be altered by the treating oncologist due to considerations like side effects and quality of life for the patient, where doses may be skipped, dosages decreased, and/or supportive medications prescribed. However, these changes are made only with a limited knowledge of their effects on the treatment outcome for any given patient. 

As patients present with varying physiologies and sensitivities, the one-size-fits all approach is clearly not optimal for all patients. 
Mathematical models of tumor growth and treatment response can help us predict therapeutic efficacy accounting for each patient's specific tumor dynamics and, potentially, select the best therapeutic regimen for each individual patient. 
For instance, Jarrett \emph{et al.} \cite{Jarrett2020} used the model in Eq.~\eqref{eq:ct} to investigate alternative dosing regimens of cytotoxic therapies for breast cancer \emph{in silico}. 
First, the model was parameterized patient-wise using MRI data collected prior to the start of therapy and after one drug cycle as indicated in Section~\ref{sec:ct}. The resulting personalized model was then simulated to the time of completion of the prescribed therapeutic regimen, and the simulated tumor growth was compared to the actual tumor response measured by MRI for each patient. The model predictions were found to be highly correlated with actual tumor response ($N = 13$, CCC$ > 0.90$, $p < 0.01$ for total cellularity, total volume, and longest axis), so the model was considered valid to reproduce the effects of chemotherapy in breast cancer.
Their validated model was then used to explore alternative therapeutic regimens, which were defined patient-wise by fixing the total dose prescribed in standard regimen and varying frequency and dosage.
The authors indicated that an additional 0-46\% reduction (median=17\%) in total cellularity may have been achievable across the patient cohort ($N = 13$) compared to the standard chemotherapeutic regimens that the patients were prescribed. The dosing regimens that the model predicted to reduce/control each tumor were also found to significantly outperform standard regimens for tumor control ($p < 0.001$), thereby supporting the claim that standard regimens may not be the most effective for every patient.

\subsection{Optimal control theory for personalized treatment planning} \label{sec:8.3}
Consider a dynamical system involving a set of variables $u(t)$ and controls $z(t)$, which are functions describing external forces that can alter the system dynamics.
\emph{Optimal control theory} (OCT) was developed to determine the solution of the system that achieves a particular outcome by adequately adjusting the controls.
The mathematical formulation of the optimal control problem consists of minimizing or maximizing an objective functional $J(u,z)$. Thus, given a particular dynamical system over a certain time interval $[0,T]$, applying OCT largely consists of determining the objective functional, problem-specific constraints, and a method for solving the OCT problem. 
The general form of the objective functional for OCT is
\begin{equation} 
J(u,z) = \Phi[u(t_j),z(t_j),t_j]+\int_{t_0}^{t_f}L(u(t),z(t),t)dt, 
\label{eqn:ObjFunc}
\end{equation}
where $\Phi[u(t_j),z(t_j),t_j]$ includes target values of the variables and the controls at specific times $\{t_k\}_{k=1,\ldots,n_t}$, while $L(u(t),z(t),t)$ accounts for the target dynamics of the variables and the controls over $[0,T]$. $\Phi[u(t_j),$ $z(t_j),t_j]$ has several names in OCT literature, including  \emph{endpoint cost} in minimization problems and \emph{terminal payoff} in maximization problems. 
The formulation of the objective function can also be divided into three canonical types: \emph{endpoint control}, which only includes $\Phi[u(t_j),z(t_j),t_j]$; \emph{bang-bang control}, which only features a linear $L(u(t),z(t),t)$; and \emph{continuous control}, which only has a quadratic $L(u(t),z(t),t)$. The objective function can also be constructed by combining these canonical types.
The term \emph{bang-bang} refers to the usual dynamics of optimal control $z(t)$ for this type of functional, which switches between the maximum admissible value and $z(t)=0$ (i.e., no effect).
Additionally, the quadratic term in the continuous control is not usually motivated by problem-dependent phenomena, but it ensures that the optimization problem is \emph{convex}. Hence, the optimal control problem has key mathematical features, including the existence of a global minimum.

In the context of cancer, we can apply OCT to obtain optimal treatment strategies by using a mathematical model to simulate tumor growth and therapeutic response as a dynamical system, setting the treatment as a control, and selecting clinically-relevant treatment outcomes in the objective functional.
Here we will briefly discuss the formulation of the OCT problem using a simplification of the PDE model of breast cancer chemotherapy in Eq.~\eqref{eq:ct}, whereby we directly model the total number of tumor cells $n(t)$ using the ordinary differential equation (ODE) \cite{Yin2019,Johnson2020}:
\begin{equation}\label{ode_nat}
\frac{dn(t)}{dt}=kn(t)\left(1-\frac{n(t)}{\theta}\right)-\alpha z(t)n(t)
\end{equation}
where $k$ describes global tumor cell proliferation, $\theta$ is the tissue carrying capacity, $\alpha$ models chemotherapy efficacy, and $z(t)$ represents the dynamics of the concentration of drug(s) in the plasma, which is used to derive the initial concentration of drug in the tissue $C_{drug}(\vec{x},t)$ in Eq.~\eqref{eq:ct} \cite{Jarrett2018a}. We chose this ODE model for the sake of simplicity and because ODE models have been commonly used in the cancer OCT literature \cite{Jarrett2020a, Lenhart2007,Schaettler2015, Anita2011, Swan1990}. However, the following ideas could also be applied to the PDE model in Eq.~\eqref{eq:ct} by defining $n(t)=\int_{\Omega}N(\vec{x},t)d\vec{x}$ (see \cite{Almeida2019}).

\begin{figure}[!t]
\includegraphics[width=\textwidth]{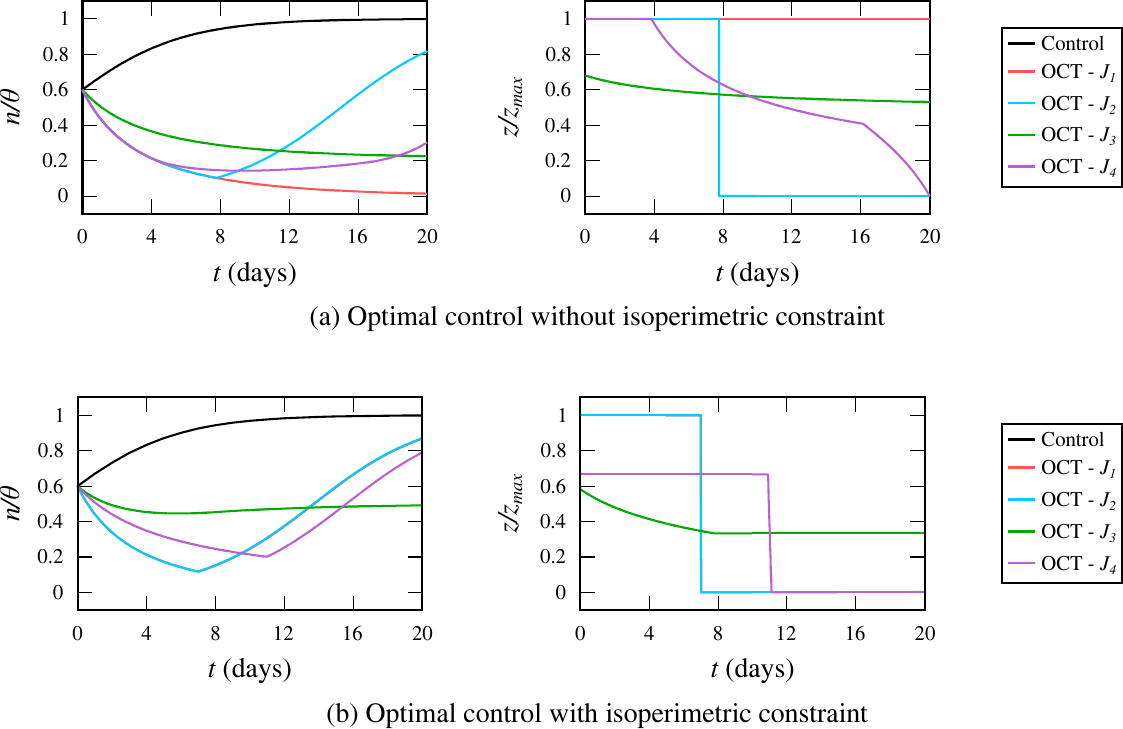}
\caption{Optimal control problem solutions for the ODE model of cancer chemotherapy in Eq.~\eqref{ode_nat}.
(a) Without an isoperimetric constraint, the optimal control problem using $J_1$ yields $z(t)=z_{max}$, which would probably induce acute toxicity in patients. The optimal $z(t)$ for the bang-bang control using $J_2$ produces the characteristic stepwise solution, while the optimal $z(t)$ obtained with the continuous controls in $J_3$ and $J_4$ show a continuous decay. The drug regimens calculated with $J_1$, $J_3$ and $J_4$ successfully control tumor growth, while for $J_2$ we observe regrowth once $z(t)$ drops to zero. 
(b) If we impose the isoperimetric constraint, the optimal $z(t)$ for $J_1$ and $J_2$ coincide and produce a similar regimen to the bang-bang solution without isoperimetric constraint. The optimal $z(t)$ obtained with the continuous controls in $J_3$ and $J_4$ now take lower concentration values, which translate into a limited tumor control. Balancing the maximum drug delivered during treatment, the maximum reduction in tumor burden, and the relative weights in the formulations of $J_2$, $J_3$, and $J_4$ is a challenge in OCT applications in cancer, which may render very different results.  }
\label{fig:optcon}
\end{figure}

A primary goal of OCT problems for cancer treatment is to minimize the tumor burden only at the completion of the therapy, which can be formulated by the endpoint control functional
\begin{equation}
J_1(n)=n(t_f).
\label{eq:endpointcost}
\end{equation}
This optimal control problem requires additional constraints for the therapeutic regimen $z(t)$. For example, these can limit the maximum dose by setting $z(t)<z_{max}$ and/or the total maximum dose by imposing $\int_{0}^{T}z(t)dt < z_{tot}$, which is termed an \emph{isoperimetric constraint} (see Fig.~\ref{fig:optcon}).
The limits $z_{max}$ and $z_{tot}$ are drug-dependent and may be patient-specific (e.g., quality of life, comorbidities). However, the optimization of $J_1(n)$ is insensitive to the dynamics of the drug concentration; i.e., it does not formally adapt the drug regimen to the tumor burden.  Another limitation of endpoint control formulations is that they do not consider the potential growth of the tumor during therapy, which may be relevant to select actionable therapeutic regimens depending on the type of cancer.

To address this limitation, we can extend the objective functional in Eq.~\eqref{eq:endpointcost} with a bang-bang term accounting for $z(t)$:
\begin{equation}
J_2(u,z)=w_1 n(t_f)+w_2\int_{0}^{T}z(t)dt.
\label{eq:bangbang}
\end{equation}
where $w_1$ and $w_2$ are problem specific weights that can be included in the objective function to give greater or lesser importance to the different terms in the objective functional during optimization. For example, if a particular treatment has significant adverse side effects, $w_2 > w_1$  may be enforced to focus the problem on minimizing the total drug dose. Similarly, an objective function that includes the continuous control of the therapy can be written as
\begin{equation}
J_3(u,z)=w_1 n(t_f)+\frac{w_2}{2}\int_{0}^{T}(z(t))^2dt.
\label{eq:continuousdrug}
\end{equation}
Furthermore, the objective function can also account for the tumor growth over the entire treatment period:
\begin{equation}
J_4(u,z)=\frac{1}{2}\int_{0}^{T}w_1(n(t))^2+w_2(z(t))^2dt.
\label{eq:continuousboth}
\end{equation}

The main difference between the optimal treatment solution found for objective function $J_2$ versus $J_3$ and $J_4$ is that $J_2$ assumes that a constant dose is given over a certain interval, while $J_3$ and $J_4$ assume that the drug concentration $z(t)$ can change over time. These differences are shown in Fig.~\ref{fig:optcon}.
In principle, an on-off (bang-bang) control would be more clinically relevant than the continuous control, as patients are not usually treated over time in a continuous manner. However, this type of control would result in a sudden drop in $z(t)$ after the conclusion of each drug cycle (see Fig.~\ref{fig:optcon}), whereas in reality $z(t)$ decays with an approximately exponential trend. With the introduction of take-home infusion pumps for chemotherapy \cite{Zahnd1999}, evaluating continuous control may be a more plausible avenue of investigation. 
However, the optimal $z(t)$ obtained under continuous control may also be unachievable with the current drugs for cancer treatment (e.g., due to incompatible pharmacokinetics).
Ultimately, the optimal drug concentrations $z(t)$ obtained with these functionals can render valuable information to guide the design of clinically-feasible optimal therapeutic strategies. Another future goal could even be the synthesis of new drugs or the adaptation of current drug compounds to match target dynamics emanating from the combination of OCT, cancer modeling, and pharmacodynamics \cite{Iyengar2012,Shi2017,Luepfert2005}.

Finally, optimal control problems using $J_2$, $J_3$, and $J_4$ may include clinically-relevant constraints to further focus the solution, including the limitations to $z(t)$ discussed for $J_1$. For instance, we can limit the tumor burden at any time by imposing $n(t)<n_r$, where $n_r$ is an arbitrary threshold. Additional constraints can also limit the frequency $f$ of doses, for example, by setting $f \leq T/n_d$, where $n_d$ is the maximum number of doses per treatment period.
Larger systems of equations that may account for the healthy and immune cell populations may also require additional constraints and/or incorporate these other variables into the optimal control problem itself. Beyond biological concerns are the logistical, monetary, and psychological costs that may also be considered for an optimal regimen.
The reader is referred to \cite{Jarrett2020a, Lenhart2007,Schaettler2015, Anita2011, Swan1990} for further details on applying OCT to cancer models.

\section{Barriers to success}\label{sec:barriers}

One of the principal issues preventing the successful translation of image-based computational modeling technologies to routine clinical practice is access to proper data. 
Calibration and validation of image-based cancer models require individual quantitative imaging data at multiple time points during the course of surveillance or treatment.
The acquisition of frequent and rich imaging datasets might be feasible in preclinical settings, thereby enabling the realization of controlled studies to thoroughly assess model validity. 
However, such controlled studies are extremely challenging in a clinical scenario for two key reasons. First, the timely and effective treatment of the patient and their quality of life are the utmost priorities. Thus, a controlled validation study would require the acquisition of minimal datasets to ensure a reliable calibration of the model, while ensuring that data collection is not a burden for patients and that those in the model-informed arm will be closely monitored in case immediate action is warranted during the course of the controlled study.
Second, quantitative imaging data are not routinely collected in the current clinical protocols. 
Standard-of-care imaging primarily focuses on delineating tumor boundaries for staging and planning interventions (e.g., biopsy, surgery, and radiotherapy). However, anatomical imaging fundamentally limits modeling approaches as it does not quantitatively characterize the unique heterogeneous nature of each patient's tumor. 
In addition, standard-of-care images are often acquired only before and following the completion of therapy. This is partly due to financial constraints, hospital workflow, scanner availability, and patient burden. 
Thus, carefully designed studies are needed to determine a clinically-feasible strategy to collect sufficient quantitative imaging data enabling an accurate parameterization of predictive models of cancer.
For instance, abbreviated imaging protocols may fit quantitative, research-focused scans into routine clinical visits alleviating the need for separate research scans.
Additionally, many quantitative imaging data types can now be acquired in the community setting (i.e., away from major hospitals or oncology centers), which may be more convenient for patients \cite{Sorace2018}.

From a modeling viewpoint, a central challenge is balancing model complexity, a rational use of computational resources, data requirements to ensure an accurate calibration, and predictive accuracy.
The dynamics of cancer growth and treatment response is extremely complex, involving a multitude of biophysical processes interacting at various spatial and temporal scales \cite{Marusyk2010}.
Mathematical models of cancer are built upon a series of relevant biophysical phenomena, whose selection, formulation, and calibration ultimately determines the predictive power of the model \cite{Yankeelov2013,Yankeelov2015,Rockne2019}. 
The accurate modeling of some of these phenomena (e.g., angiogenesis \cite{Vilanova2017}) would require complex equations at multiple scales, a large number of parameters requiring extensive spatiotemporal data, and advanced numerical methods that may incur in a large computational cost.
However, such models would be incompatible with the constraints on quantitative imaging data availability and patient care during the controlled validation studies discussed above.

Thus, clinically-oriented image-based models usually require conservative modeling assumptions, which enables the description of cancer growth and treatment response at organ scale using simple modeling paradigms and involving a minimal set of parameters whose calibration is feasible with existing quantitative imaging techniques \cite{Yankeelov2013}.
However, these model assumptions may incur substantial errors that ultimately limit the predictive power of the model.
Moreover, some biophysical model fields (e.g., tumor cell density) are not directly observed. Instead, the observed imaging data must be preprocessed to create a proxy to the biophysical observables. Such \emph{pseudo-correspondences} introduce additional uncertainties and can significantly affect the reconstruction results~\cite{Gholami:2016a,Swanson:2008a,Le:2017a,Harpold:2007a}.
Likewise, the phenomenological nature of macroscopic cancer models introduces parameters without direct biological counterparts (e.g., tumor cell diffusivity, drug efficacy), which further complicates model assessment and validation.

Nevertheless, the future development of image-based cancer models requires the initial validation of simpler models because this enables the identification of key improvements in model formulation to refine predictive accuracy. In addition, the success of simpler models also provides justification to collect further data to calibrate more complex biophysical mechanisms in subsequent model extensions.
For example, this has been the process behind the extension of the logistic model in Eq.~\eqref{eq:log} to the reaction-diffusion model in Eq.~\eqref{eq:rd}, then to a mechanically-coupled  model in Section~\ref{sec:mech}, and later to a vasculature-coupled or drug-informed model in Sections~\ref{sec:vasc} and \ref{sec:ct}, respectively.

Biophysical inversion is a promising strategy to calibrate predictive models of cancer, but also presents several challenges. Complex, typically nonlinear and time-dependent, PDE-based models often result in ill-conditioned and non-convex optimization problems, which require sophisticated numerical algorithms to stabilize the inversion, such as multiresolution continuation, parameter continuation, and regularization schemes \cite{Scheufele:2020,Scheufele:2019a,Gholami:2016a,Subramanian20IP}. 
Data scarcity can exacerbate the ill-posedness.
Mitigating this issue can entail imposing additional modeling assumptions and regularization strategies \cite{Subramanian20IP}. 
The noise arising from various sources in imaging data also complicates model inversion. Further modeling priors and structure-exploiting algorithms can help mitigate some of these issues. Other mathematical considerations such as the choice of mismatch function (e.g., $L^2$ loss, cross-entropy loss) and regularization models can further complicate the inversion.
Inversion methods may also require several forward model evaluations, so specialized solvers are needed to prevent prohibitive computational costs.

OPAL is an attractive methodology to decide potential model extensions by comprehensively assessing the improvement in model predictions against the increase in model complexity.
In particular, OPAL can guide the modeler to select the \emph{best} valid model representing a certain experimental or clinical setting while accounting for uncertainties in both data and model parameters.
However, OPAL may also face certain challenges. 
For example, computationally expensive models might require a more efficient approach for sensitivity analysis than the Sobol method, such as the elementary effects method or metamodels \cite{saltelli2008global}. Another major difficulty in model selection is that every model in the initial set of models might be invalid (i.e., they do not satisfy the validation criteria), which would require to extend such set with further models.

Finally, application of OCT to mathematical models of tumor growth and treatment response is a promising strategy for the optimization of therapeutic regimens \textsl{in silico} \cite{Jarrett2020}. However, the reliability and plausibility of solutions generated by OCT methods depend on several factors including the validity of the model, the accurate definition of the objective functional, the uncertainty in the parameters and data, the application of clinically-relevant constraints, and the accuracy in solving the OCT problem itself.
Additionally, implementation of OCT approaches within the clinical trial system is even more complicated than model validation. Beyond the challenges on data availability, model assumptions, and biophysical inversion described above, a controlled clinical study to validate OCT-generated therapies would involve the test of novel computationally-derived regimen protocols in patients. This requires a close monitoring of treatment response, toxicity, and patient well-being, which can be extremely complex to balance and maintain during such clinical study. Thus, robust preclinical evidence showing the advantages of OCT for the design of therapeutic regimens is required before advancing to clinical scenarios.

\section{Conclusion}\label{sec:conc}

Integrating quantitative data obtained from biomedical imaging with mechanism-based mathematical modeling represents a significant departure from current para-digms in cancer biology and oncology.  More specifically, this approach is fundamentally different from the current trend in modeling which emphasizes applying the methods of artificial intelligence to extremely large data sets.  However, statistical inference - though enormously powerful - relies on properties of large populations that can frequently obscure important characteristics (or conditions) that are specific to individual patients and may drive the development of their disease or their response to therapy.  The high-consequence decisions present in clinical oncology simply must be based on more than data analytics. These decisions must incorporate biophysical processes within a rigorous, mathematical framework that can be calibrated with patient-specific data to make patient-specific predictions.  The transformation from population-based care to patient-based care is inevitable, and the intimate integration of quantitative imaging, mechanism-based mathematical modeling, and efficient computational methods enabling  precise \emph{in silico} tumor forecasts is a very promising avenue to achieve this important goal.

\section*{Acknowledgements}
GL was partially supported by a Peter O'Donnell Jr. Postdoctoral Fellowship from the Oden Institute for Computational Engineering and Sciences at The University of Texas at Austin and acknowledges funding from the European Union's Horizon 2020 research and innovation programme under the Marie Sk\l{}odowska-Curie grant agreement No. 838786.
EABFL is supported by the Oncological Data and Computational Sciences collaboration between the Oden Institute for Computational Engineering and Sciences, The University of Texas MD Anderson Cancer Center, and Texas Advanced Computing Center.
TEY is a CPRIT Scholar in Cancer Research and acknowledges funding from CPRIT RR160005, NCI U24CA226110, NCI R01CA235800, and NCI U01CA253540.



\begin{small}

\end{small}

\end{document}